\documentclass[10pt,preprint,compsoc]{IEEEtran}

\usepackage{ifpdf}        
\usepackage{amsmath}      
\usepackage{algorithmic}  
\usepackage{array}        
\usepackage{url}          

\usepackage{breakurl}
\usepackage[breaklinks]{hyperref}

\ifpdf
\else
\fi

\ifCLASSOPTIONcompsoc
  \usepackage[nocompress]{cite}
\else
  \usepackage{cite}
\fi

\ifCLASSINFOpdf
   \usepackage[pdftex]{graphicx}
   \graphicspath{{figures/}{pictures/}{images/}{./}}  
   \DeclareGraphicsExtensions{.pdf,.png,.jpg,.jpeg}   
\else
   \usepackage[dvips]{graphicx}
   \graphicspath{{figures/}{pictures/}{images/}{./}}  
   \DeclareGraphicsExtensions{.eps}                   
\fi

\ifCLASSOPTIONcompsoc
 \usepackage[caption=false,font=footnotesize,labelfont=sf,textfont=sf]{subfig}
\else
 \usepackage[caption=false,font=footnotesize]{subfig}
\fi

\ifCLASSOPTIONcaptionsoff
 \usepackage[nomarkers]{endfloat}
\let\MYoriglatexcaption\caption
\renewcommand{\caption}[2][\relax]{\MYoriglatexcaption[#2]{#2}}
\fi

\usepackage{amsmath}
\usepackage{csquotes}
\usepackage{enumitem}

\usepackage{booktabs}
\usepackage{multirow}
\usepackage{graphicx}
\usepackage{soul}

\usepackage{tabularx}
\usepackage{tabulary}

\usepackage{subfig}

\usepackage{csquotes}
\usepackage[table]{xcolor}

\usepackage{color}
\usepackage{xcolor}

\usepackage{makecell}

\usepackage{fontawesome}

\usepackage[normalem]{ulem}
\definecolor{mred}{rgb}{.80,.12,.30}
\definecolor{MRED}{rgb}{.80,.12,.30}
\definecolor{grey}{rgb}{0.5,0.5,0.5}
\definecolor{purple}{rgb}{.75,0,.85}
\definecolor{pistachio}{rgb}{0.58, 0.77, 0.45}

\newif\ifnotes
\notestrue

\let\origcite\cite
\renewcommand{\cite}[1]{\ifnotes\mbox{\origcite{#1}}\else \origcite{#1}\fi}

\definecolor{t10h_1_dblue}{rgb}{0.0, 0.2, 0.6}
\definecolor{t10l_1_dblue}{rgb}{.68, .78, .91}
\definecolor{darkmidnightblue}{rgb}{0.0, 0.2, 0.4}
\definecolor{midnightblue}{rgb}{0.1, 0.1, 0.44}
\definecolor{darkpowderblue}{rgb}{0.0, 0.2, 0.6}

\definecolor{t10h_2_orange}{rgb}{1, .5, .05}
\definecolor{t10l_2_orange}{rgb}{1, .73, .47}

\definecolor{t10h_3_dgreen}{rgb}{.17, .63, .17}
\definecolor{t10l_3_dgreen}{rgb}{.6, .87, .54}

\definecolor{t10h_4_red}{rgb}{0.55, 0.0, 0.0}
\definecolor{t10l_4_red}{rgb}{1, .6, .59}
\definecolor{darkred}{rgb}{0.55, 0.0, 0.0}
\DeclareRobustCommand{\hlgfour}[1]{{\sethlcolor{t10l_4_red}\textbf{\textcolor{t10h_4_red}{\hl{#1}}}}}
\definecolor{t10h_5_purple}{rgb}{0.4, 0.19, 0.28}
\definecolor{t10l_5_purple}{rgb}{0.86, 0.82, 1.0}
\definecolor{lightmauve}{rgb}{0.86, 0.82, 1.0}
\definecolor{oldmauve}{rgb}{0.4, 0.19, 0.28}

\definecolor{t10h_6_brown}{rgb}{.55,.34,.29}
\definecolor{t10l_6_brown}{rgb}{.77, .61, .58}

\definecolor{t10h_7_pink}{rgb}{.89,.47,.76}
\definecolor{t10l_7_pink}{rgb}{.97, .71, .82}

\definecolor{darkolivegreen}{rgb}{0.33, 0.42, 0.18}
\definecolor{olivedrab7}{rgb}{0.33, 0.42, 0.18}
\definecolor{t10h_9_ggreen}{rgb}{0.33, 0.42, 0.18}
\definecolor{t10l_9_ggreen}{rgb}{.86, .86, .55}
\DeclareRobustCommand{\hlgeight}[1]{{\sethlcolor{t10l_9_ggreen}\textbf{\textcolor{t10h_9_ggreen}{\hl{#1}}}}}

\definecolor{t10h_10_lblue}{rgb}{0.24, 0.56, 0.87}
\definecolor{t10l_10_lblue}{rgb}{.62, .85, .9}
\definecolor{tuftsblue}{rgb}{0.24, 0.56, 0.87}
\DeclareRobustCommand{\hlgnine}[1]{{\sethlcolor{t10l_10_lblue}\textbf{\textcolor{t10h_10_lblue}{\hl{#1}}}}}

\DeclareRobustCommand{\hlgVis}[1]{{\sethlcolor{t10l_10_lblue}\textbf{\textcolor{t10h_10_lblue}{\hl{#1}}}}}

\DeclareRobustCommand{\hlgLimit}[1]{{\sethlcolor{t10l_4_red}\textbf{\textcolor{t10h_4_red}{\hl{#1}}}}}

\DeclareRobustCommand{\hlgContext}[1]{{\sethlcolor{t10l_9_ggreen}\textbf{\textcolor{t10h_9_ggreen}{\hl{#1}}}}}

\usepackage{wrapfig}

\begin{document}


%
\title{Are Metrics Enough? Guidelines for Communicating and Visualizing Predictive Models to Subject Matter Experts}
%
%
\author{Ashley Suh, Gabriel Appleby, Erik W. Anderson, Luca Finelli, Remco Chang, Dylan Cashman

\IEEEcompsocitemizethanks{
\IEEEcompsocthanksitem
1077-2626 $\copyright$ 2023 IEEE. Personal use is permitted, but republication/redistribution requires IEEE permission.
See \url{https://www.ieee.org/publications/rights/index.html} for more information.
\IEEEcompsocthanksitem
Ashley Suh, Gabriel Appleby, and Remco Chang are with Tufts University.  E-mail: \{ashley.suh, gabriel.appleby, remco.chang\}@tufts.edu
\IEEEcompsocthanksitem
Erik W. Anderson, Luca Finelli, and Dylan Cashman are with Novartis Pharmaceuticals Corporation, Data Science and AI. E-mail: \{erik.anderson, luca.finelli, dylan.cashman\}@novartis.com.}}
\markboth{\textit{$\copyright$~2023 IEEE. This is the author’s version of the article. The final version of this record is available at: \href{https://ieeexplore.ieee.org/document/10077087}{10.1109/TVCG.2023.3259341}}}%
{}
\IEEEtitleabstractindextext{%
\begin{abstract}
Presenting a predictive model's performance is a communication bottleneck that threatens collaborations between data scientists and subject matter experts.  Accuracy and error metrics alone fail to tell the whole story of a model -- its risks, strengths, and limitations -- making it difficult for subject matter experts to feel confident in their decision to use a model.  As a result, models may fail in unexpected ways or go entirely unused, as subject matter experts disregard poorly presented models in favor of familiar, yet arguably substandard methods.  In this paper, we describe an iterative study conducted with both subject matter experts and data scientists to understand the gaps in communication between these two groups.  We find that, while the two groups share common goals of understanding the data and predictions of the model, friction can stem from unfamiliar terms, metrics, and visualizations -- limiting the transfer of knowledge to SMEs and discouraging clarifying questions being asked during presentations.  Based on our findings, we derive a set of communication guidelines that use visualization as a common medium for communicating the strengths and weaknesses of a model.  We provide a demonstration of our guidelines in a regression modeling scenario and elicit feedback on their use from subject matter experts.  From our demonstration, subject matter experts were more comfortable discussing a model's performance, more aware of the trade-offs for the presented model, and better equipped to assess the model's risks -- ultimately informing and contextualizing the model's use beyond text and numbers.
\end{abstract}

\begin{IEEEkeywords}
Visualization techniques and
methodologies, Human factors, Modeling and prediction
\end{IEEEkeywords}}

\maketitle

\IEEEdisplaynontitleabstractindextext


\IEEEraisesectionheading{
  \section{Introduction}
  \label{sec:introduction}
}
\IEEEPARstart{T}{}he widespread use of artificial intelligence (AI) has reached far beyond academia. In a recent study with over 70 Fortune 1000 companies, the percentage of firms investing at least \$50 million in AI increased by over 60\% from 2018 to 2020\cite{algorithmia2021enterprise}. 
Surprisingly though, despite increased investments in AI, only \textbf{15\%} of these firms reported that they had \textit{actually deployed} AI capabilities into widespread production. Similar findings were echoed by Ransbotham et al. in an MIT SMR report: 7 out of 10 companies reported minimal to no impact from their organization's AI use, with only 10\% of companies seeing significant financial benefits from implementing AI solutions\cite{ransbotham2020expanding}.

The potential benefits of AI and machine learning (ML) models have driven great interest across a variety of domains, such as healthcare\cite{jiang2017artificial}, biology\cite{webb2018deep}, and commercial industries\cite{furman2019ai}.  
While companies invest millions into AI and ML production, it is evident that a gap still exists between the promise and the actual execution of these models in practice\cite{algorithmia2021enterprise}. 
Data science interviews\footnote{\url{https://hbr.org/2019/01/data-science-and-the-art-of-persuasion}} and qualitative studies\cite{passi2018trust, hong2020human, berinato2019data} consistently cite \textit{strained communication} between data scientists and subject matter experts as a top barrier to model development and deployment. 

Communication between members of different disciplines and expertise is a difficult task that is often facilitated by visualization\cite{smiciklas2012power, bottinger2020reflections, kosara2013storytelling, ma2011scientific}. However, while visualization has acted as a powerful communication medium for broad audiences\cite{segel2010narrative}, most research in visualization for AI/ML has focused on helping data scientists develop, debug, and improve their models\cite{sacha2018vis4ml}. We believe that a more fundamental visualization question needs to be answered for AI and ML development: how can we best communicate the strengths and weaknesses of a predictive model's performance to subject matter experts, who ultimately decide what to do with it? 

To investigate how visualization can mitigate issues in model communication -- which ultimately lead to downstream problems with deployment -- we conducted two interview studies with data science practitioners who develop models, and the subject matter experts (SMEs) that must work with the outputs of those models. 
From a thematic analysis in which we iteratively coded and categorized our first round of interviews, we present the most prevalent themes related to the challenges and unaddressed communication needs of SMEs in model collaborations.

We find that, while data scientists and SMEs are often concerned about similar issues (\emph{e.g.}, strengths and weaknesses of a model, edge cases, etc.), each group approaches these issues drawing upon very different sets of language, context, and experiences. SMEs reported feeling overwhelmed by data science presentations, discouraged from asking clarifying questions, and skeptical about whether a predictive model actually fits their needs. Consequently, this lack of a common framing results in misunderstandings, miscalculations of a model's risks, and in some cases the failure of a model's adoption.  While data visualization was cited as a tool employed by data scientists to better communicate model performance results, our interviews revealed that many commonplace predictive model visualizations (\emph{e.g.}, residual plots) may result in greater confusion by SMEs, particularly when they are not already familiar with technical data science concepts or visualization methods. 

To combat this confusion, data scientists diverted to presenting performance metrics, such as explained variance or mean squared error, to illustrate their model's performance in a presentation.  However, we observe that the presentation of these metrics alone is insufficient for SMEs who are responsible for making decisions and recommendations based on the model's outputs. We suggest that more easily interpretable visualization methods, such as the use of annotations and illustrative grounded use cases, can more effectively communicate a model's performance to SMEs beyond error metrics. As a result, these methods can provide a clearer picture of the model's impact and allow SMEs to better understand and assess its outcomes.

From our interview findings, we distill a set of model communication guidelines for data scientists to consider when presenting the complexities of a model's performance. We then demonstrate an application of our guidelines in a data science presentation with four visualizations, and elicit feedback from SMEs. We find that the use of visualization enabled SMEs to have ``\textit{quick}'' and ``\textit{instant}'' insights beyond text and numbers. Furthermore, when presented on a model's performance using our guidelines, SMEs felt more comfortable asking questions, better equipped to understand the model's strengths and weaknesses, and ultimately empowered to use their domain expertise to make a more calculated interpretation of the model.  

In summary, our major contributions are as follows: 

\begin{itemize}[leftmargin=*,topsep=2pt, partopsep=0pt,itemsep=1pt,parsep=1pt]
    \item Two interview studies with data scientists and SMEs who regularly collaborate on predictive models.
    \item Guidelines that emerge from the analysis of our interviews and review of related literature.
    \item Example slides and visualizations, as well as an open-source Observable notebook\footnote{\url{https://observablehq.com/@ashleysuh/model-comm-vis}} used to generate those visualizations, so that our guidelines are easy to implement for data science presentations.  
\end{itemize}

\section{Related Work}
\label{sec:related}

For the remainder of this work, we broadly refer to any practitioner who works on and builds predictive models as a \textit{data scientist}. We refer to \textit{subject matter experts (SMEs)} as model consumers with primary roles and expertise outside of data science and machine learning.  
We differentiate between model consumers who take the role as `executives' (\textit{i.e.} stakeholders) and those whose primary function is as a domain scientist (SME) that works with data scientists to develop and deploy predictive models in their daily work. Although some SMEs may also be stakeholders, we make the distinction in this paper that an SME has a specialized set of knowledge for a particular scientific domain. 

\subsection{Visualization for Model Interpretability}
\label{sec:related:visual_comm}
Effective presentation of a predictive model's performance to domain scientists, SMEs, and other stakeholders is an active topic of research in visualization.  
Researchers studying explainable AI seek to help users interpret and explain the inferences of AI models by visualizing the internal workings of those models\cite{mohseni2018survey, lipton2018mythos, weerts2019human, doshi2017towards}.  Metrics and principles are posed for explainable AI\cite{rudin2021interpretable, hoffman2018metrics}, guidelines for defining \textit{interpretability} are suggested\cite{davis2020measure, yang2020visual}, and visual analytic tools can enhance machine learning and AI transparency\cite{sacha2018vis4ml, chatzimparmpas2020state, hohman2019telegam, krause2017workflow}. Hohman et al. conducted an interview study with machine learning experts to understand how interactive interfaces could better support model interpretation for data scientists\cite{hohman2019gamut}. 
While many of these works are relevant to the research presented here, the concepts of explainability are discussed at too low of a level. Consequently, a majority of ML interpretability support is targeted towards supporting experts in debugging and improving models (\emph{e.g.},\cite{sacha2018vis4ml}) and not necessarily in communicating the limitations or weaknesses of models to the stakeholders -- or consumers -- of ML\cite{shah2019making, nittas2023beyond}.

From the results of our first interview study (Sections~\ref{sec:interviews} and~\ref{sec:findings}), we found that solutions were needed to facilitate model communication between data scientists and SMEs at a higher level, particularly when SMEs are responsible for adopting and monitoring those models in practice.
To this end, we identify what data scientists and SMEs each find most valuable in the presentation of a model's outcome, and distill common barriers SMEs face when receiving and interpreting presentations on model performance. 
From these findings, we propose visualization guidelines (Section~\ref{sec:guidelines}) that practitioners can use when communicating models.  

\subsection{Data Science \& ML in Practice}
\label{sec:related:ml_in_practice}
The desire and demand for AI/ML at organizations outside of big tech is well documented\cite{algorithmia2021enterprise, hopkins2021machine}.
Consequently, there has been a sharp increase in the use (and subsequent risk) of advanced analytics in nontraditional ML domains (\emph{e.g.}, for healthcare and childcare) \cite{evans2000pharmacovigilance, almenoff2007innovations, zytek2021sibyl}.
However, when introducing AI/ML techniques in practice, it is also necessary to investigate the risks, limitations, edge cases, and weaknesses of a predictive model before its deployment\cite{evans2000pharmacovigilance, almenoff2007innovations}. 
Although strides are continuously being made to improve the transparency of ML models, empirical research shows these techniques are rarely deployed in high stakes domains\cite{harvey2002analysis, cresswell2013organizational, emanuel2019artificial, shilo2020axes}. Close to our work is the recent design study by Zytek et al.\cite{zytek2021sibyl}, in which the authors collaborate with SMEs that use ML techniques in child welfare. 
In Zytek et al.'s work, an iterative design process combines ML practitioners and childcare experts to identify key challenges in their workflow. The final result is a visual analytics tool to help alleviate interpretability challenges for the domain experts. 
In contrast to Zytek et al.'s work and similar characterizations of ML interpretability challenges, we instead focus on  communication bottlenecks between data scientists and SMEs that prevent model assessment and deployment. 

Challenges in ML collaborations directly affect the eventual production of AI/ML solutions, particularly when the model's performance can not be well understood. Seneviratne et al.\ argue that work is needed to bridge the implementation gap of machine learning by merging ML algorithms into the ‘socio-technical’ milieu of the organization\cite{seneviratne2020bridging}. 
Similar work describes widespread systematic issues in the adaption of predictive models at healthcare organizations, citing the mismatch between stakeholders and their understanding of technological innovations\cite{cresswell2013organizational}. 
Shah et al.\ suggests that the utility of ML algorithms could be better demonstrated in practice if the end-users could better assess the performance of a predictive model without relying on standard performance metrics\cite{shah2019making}. In this work, we intentionally study how the performance of a predictive model can be effectively communicated to SMEs through visualization. Visualization and visual communication is often deployed to bridge communication and interpretability gaps, particularly when an audience may not have a similarly technical background\cite{bottinger2020reflections, kosara2013storytelling, smiciklas2012power}. 
This reasoning has been instrumental in guiding us to improve the accessibility of predictive models through the careful application of visualization for data science and SME communication.

\subsection{Characterizing ML Collaborations}
\label{sec:related:ml_workflows}
Across enterprise surveys that highlight common challenges for AI/ML collaborations, communication is continuously cited as the most difficult to overcome\cite{algorithmia2021enterprise}. 
Research in ML, visual analytics, and human factors have attempted to characterize the precise nature of challenges surrounding collaborations between ML researchers and domain scientists. 
After 6 months of fieldwork with a corporate data science team, Passi and Jackson found that data science practitioners had several frustrations concerning the communication with stakeholders\cite{passi2018trust}, such as the lack of trust from stakeholders when the results or inner workings of an ML model were being discussed.
These challenges exist in other sectors outside of data science as well. Hopkins and Booth conducted an interview study with stakeholders from organizations outside of `Big Tech' to understand how resource constraints challenge ML workflows and development\cite{hopkins2021machine}. The authors similarly found that language barriers between ML practitioners and stakeholders prevented trust and deployment of models and data science results. 

Language barriers between data scientists and SMEs can occur at many stages of the ML collaboration process. 
Hong et al. conducted an interview study with ML practitioners on model interpretability and found that data scientists had difficulty assessing the prior expertise and knowledge level of SMEs when delivering results, presentations, and insights\cite{hong2020human}.
Similar in spirit to our work, Mosca et al. presented a study of client-facing data scientists to identify the common communication strategies they employ to translate the (potentially ill-defined) analysis needs of SMEs\cite{mosca2019defining}. 
In this paper, we focus primarily on how \textit{predictive models} (rather than general data science findings) can be best communicated with and to SMEs.  To this end, we interview both data scientists and SMEs to get a sense of current gaps in their practices.
Further, unlike prior work that aims to characterize and offer broad solutions to common ML collaboration challenges, we distill and validate visualization strategies that reduce the burden of presenting and communicating model performance.

In recent work that surveys previous studies on ML collaborations, Suresh et al. suggest that the end-users for ML can be characterized, and thus better understood, beyond standard \emph{expert} versus \emph{non-expert} roles\cite{suresh2021beyond}. Specifically, the authors suggest that consumers of machine learning models (\emph{e.g.}, stakeholders, SMEs) can be categorized by their expertise (formal, instrumental, or personal) and the contexts in which this expertise manifests (ML, data domain, or the general milieu). 
To situate our work with Suresh et al.'s framework, we interview \emph{subject matter experts}, not necessarily \emph{stakeholders}, with formal knowledge of a particular data domain (\emph{e.g.}, clinical safety or biology) and personal or instrumental knowledge of machine learning. Motivated by Suresh et al.'s work, we are able to investigate how communication can be improved between data scientists and those with specialized knowledge in a respective data domain (SMEs). 

\begin{table*}[ht!]
    \centering
    \renewcommand{\arraystretch}{1.1}
    \centering
    \scalebox{0.9}{

\definecolor{palesilver}{rgb}{0.9, 0.9, 0.9}

\renewcommand\theadalign{bt}
\renewcommand\theadfont{\bfseries}
\renewcommand\theadgape{\Gape[4pt]}
\renewcommand\cellgape{\Gape[4pt]}


\resizebox{\textwidth}{!}{
\begin{tabular}[t]{llllllll}
\toprule

\thead{PID} & 
\thead{Role} & 
\thead{Education} & 
\thead{Domain expertise} &
\thead{Experience with \\ data science (1-5)} &
\thead{Frequency working \\ with data (1-5)} & 
\thead{Frequency using \\ regression (1-5)} \\

\hline

PID1  & DS  &  Doctorate & Data science & (3) Familiar       & (5) Every day & (3) 1-3x/week \\
PID2  & DS  &  Doctorate & Data science & (5) Expert        & (4) 1-3x/day  & (2) 1-3x/month \\
PID3  & DS  &  Masters   & Data science & (5) Expert        & (4) 1-3x/day  & (3) 1-3x/week \\
PID4  & DS  &  Doctorate & Data science & (5) Expert        & (5) Every day & (3) 1-3x/week \\
PID5  & DS  &  Masters   & Data science & (4) Quite familiar & (5) Every day & (2) 1-3x/month \\
PID6 & DS  &  Doctorate & Data science & (5) Expert        & (3) 1-3x/week & (5) Every day \\
PID7 & DS  &  Masters   & Data science & (5) Expert        & (3) 1-3x/week & (2) 1-3x/month \\


\rowcolor{palesilver} 
PID8  & SME &  Masters & Pharmacovigilance & (4) Quite familiar     & (3) 1-3x/week & (2) 1-3x/month \\
\rowcolor{palesilver} 
PID9  & SME   & Masters   &  Pharmacovigilance & (3) Familiar           & (5) Every day   & (3) 1-3x/week \\
\rowcolor{palesilver} 
PID10  & SME   & Bachelors & Quality Assurance & (3) Familiar           & (3) 1-3x/week & (2) 1-3x/month \\
\rowcolor{palesilver} 
PID11  & SME & Masters   & Commercial & (2) Somewhat familiar  & (3) 1-3x/week & (1) Never \\
\rowcolor{palesilver} 
PID12 & SME  & Masters   & Finance & (3) Familiar           & (5) Every day   & (2) 1-3x/month\\
\rowcolor{palesilver} 
PID13 & SME  & Masters   & Quality Assurance & (4) Quite familiar     & (5) Every day   & (3) 1-3x/week \\

\hline
\end{tabular}}

}
    \vspace{3pt}
    \caption{%
    Demographics for our interviewees in Section~\ref{sec:interviews}. Participants (DS=data scientist, SME=subject matter expert) self-reported their highest education, domain expertise, level of experience with data science (1=No experience, 2=Somewhat familiar, 3=Familiar, 4=Quite familiar, 5=Expert), as well as their frequency working with data and using regression models (1=Never, 2=1-3x/month, 3=1-3x/week, 4=1-3x/day, 5=Every day).}%
    \label{tab:interview_demographics}
\end{table*}

\section{Interview Study}
\label{sec:interviews}

Our research explores how communication can be improved between data scientists and subject matter experts when their end-goal is to use a predictive model. We conducted an interview study at a major organization with two participant groups - data scientists who build and present the performance of predictive models, and subject matter experts who make decisions with these models in their daily workflow.

The goal of our interview study was to understand the types of challenges both data scientists and SMEs can experience when assessing predictive models. 
By comparing and contrasting themes in the responses of each participant group, we could then identify opportunities for visualization to bridge communication gaps between the two groups and attain a shared understanding.

During the interviews, each participant was asked the same set of questions and given the same prompts (described in Section~\ref{sec:interviews:study_design}). Participants' feedback was elicited on a hypothetical modeling scenario, followed by a discussion of their own experiences in past model collaborations.
In addition to barriers that affect model communication, we also investigated whether any visualization techniques or presentation styles had been helpful when data scientists communicate the performance of their models.

\subsection{Participants}
\label{sec:interviews:protocol}

We solicited interview participants through email recruitment within the Data and AI division of a large pharmaceutical company.
To produce visualization guidelines that could be broadly used by practitioners, we followed the 2017--2021 Kaggle Machine Learning \& Data Science Surveys\footnote{\url{https://www.kaggle.com/c/kaggle-survey-2021}} in which \textit{regression} was marked as the most common machine learning method used by practitioners. Therefore, we required all participants to have prior experience with \textit{regression models} as a common baseline in our emails.

In our email exchange, potential participants were informed that the purpose of the interview was to discuss their experiences interpreting and communicating a regression model's performance. When soliciting SMEs, we specifically asked for ``SMEs who do not build models themselves, but have prior experience using a regression model, \emph{e.g.}, looking at its predictions, deciding whether to use it, etc.''  In contrast, when recruiting data scientists, we targeted those that ``have worked with, assessed, or communicated the performance of a regression model previously.'' In total, we interviewed 7 data scientists and 6 subject matter experts in varying departments and data domains. Our participants' demographics, as well as their self-reported level of data science experience and frequency working with regression models, can be found in Table~\ref{tab:interview_demographics}.

\subsection{Protocol}
\label{sec:interviews:study_design}
\urlstyle{tt}

All of our interviews were semi-structured, lasting roughly one hour each.
Interviews were conducted virtually on Microsoft Teams with audio only. Shortly before each interview, participants were given a copy of the consent form which contained information about the study, its design, and their rights as participants. Each participant verbally consented to the study over a recording and was given an anonymous demographics survey to complete. 

Participants were shown a set of prepared slides to elicit feedback on their experience assessing and communicating the performance of regression models. Regardless of the participant's organizational role (data scientist or SME), the same set of questions and prompts were given during the interview. The slide deck used in our interviews, as well as all interview questions asked to participants, are included in our supplemental 
 and can be accessed at \url{https://github.com/TuftsVALT/ModelComm}. The three major topics discussed during our interviews were:

\urlstyle{same}

\begin{enumerate}[leftmargin=*,topsep=4pt, partopsep=0pt,itemsep=2pt,parsep=2pt]
    \item \textbf{What would you need to know about a regression model before recommending its use?}
    Participants were described a hypothetical scenario in which a recently developed regression model was being presented at their workplace. Participants then told us what they would need to know about the model to recommend its use. Additionally, we asked participants their preference for presentation styles (single slide, in-depth, or interactive), as well as the types of visuals and information they typically see in similar data science presentations.
    
    \item \textbf{How have you assessed and communicated a regression model at work previously?}
    We asked participants to reflect on a regression model that had been previously introduced to their daily work. Once the participant had a particular model in mind, they described how its performance was assessed, communicated, and scrutinized during its development and deployment.
    
    \item \textbf{How could communicating a regression model's performance with \textit{data scientists} and \textit{subject matter experts} be improved?} We asked participants to describe how their future collaborators could better communicate and give feedback on the performance of regression models. Specifically, we asked SMEs what they would like data scientists to communicate to them regarding a model's performance, and data scientists were asked what they would like SMEs to communicate to them. 
\end{enumerate}

The topics were designed to engage participants in a broad discussion on factors that influence their trust and interpretation of a regression model's performance and outputs. By discussing past experiences as well as hypothetical scenarios, we could gather responses from both participant groups to identify opportunities for improved communication. Ultimately, we use these findings (detailed in Section~\ref{sec:findings}) to inform our guidelines in Section~\ref{sec:guidelines}.

\subsection{Analysis Methodology}
\label{sec:interviews:methodology}

To identify communication challenges that threaten model collaborations between data scientists and SMEs, we conducted a thematic analysis of our interviews. 
We followed a top down approach, described by Braun and Clarke\cite{braun2006using}, in which we iteratively coded and categorized statements made by participants to discover prevalent themes.

Developing codes for our codebook was guided by literature in qualitative analysis practices\cite{macqueen1998codebook, decuir2011developing} and visualization studies\cite{alspaugh2018futzing, kandel2012enterprise}. At the first stage of our codebook, we began with a set of theory-based codes using prior work that characterizes ML workflows (see Section~\ref{sec:related:ml_workflows}). 
In particular, we started with codes we derived from work by Suresh et al.\cite{suresh2021beyond}, which includes objectives and tasks distilled from 58 papers across computer science and social sciences on characterizing the stakeholders of machine learning. 
The first complete draft of our codebook was developed over a 2-hour working session with all authors. In addition to our theory-based codes, we added codes we considered missing based on our initial observations of the interview data. 
The remainder of our coding process largely followed the procedure described by Alspaugh et al.\cite{alspaugh2018futzing}. 

We defined a participant's utterance as an individual response to each of our interview questions, such that short responses and multi-turns taken in conversation (about the same topic) were weighted fairly. At times, participants changed topics (\emph{e.g.}, recounting a previous collaboration or memory based on an interview question), which we counted as a new utterance. Each utterance was coded as a whole and could be assigned one or more codes. 
The final iteration of our codebook totaled 44 codes. Our full codebook -- including a definition of each code, inclusion and exclusion criteria, and an example via participant quotes (following codebook protocols\cite{decuir2011developing}) -- is included as supplemental. 

To identify emergent themes, we ranked codes using several metrics including frequency, entropy, as well as the differences and similarities of the distribution of those metrics by group. The goal of this process was to perform an unbiased analysis to discern critical communication challenges, needs, and opportunities most important to both data scientists and SMEs. The three highest-level themes from this process, as deliberated by authors, were:

\begin{itemize}[leftmargin=*,topsep=4pt, partopsep=0pt,itemsep=3pt,parsep=2pt]
    \item Data science presentation styles and visualizations used when illustrating model performance
    \item Understanding a model's strengths and weaknesses
    \item Gaps and mismatches in language, context, and comfort experienced by SMEs during data science collaborations
\end{itemize}

The organization of our codes into high-level themes was conducted during four working sessions with all authors. A full report of our coding procedure (including coding passes and tie-breakers) is included in our supplemental material.

\section{Interview Findings}
\label{sec:findings}

In this section, we present the collective findings from our first round of interviews with data scientists and SMEs. We center the discussion around particular sources of friction in communication between the two groups, and identify opportunities to help SMEs better understand and assess the performance of a predictive model. From our thematic analysis, we identified and grouped challenges related to: (1) presentation and visualization styles, (2) understanding the strengths and weaknesses of a model, and (3) gaps in language, context, and comfort. For each category, we highlight the most common issues that cause model collaborations to suffer. In Section~\ref{sec:guidelines}, we map these challenges (C1-C7) to our proposed model communication guidelines.

\subsection{Presentation and Visualization Styles}
\label{sec:findings:presentation}
The first theme relates to the communication and illustration of model performance through presentation and visualization styles. SMEs revealed to us that data science presentations, metrics, and visualizations are not as easily interpretable or recognizable as data scientists think: ``\textit{Data scientists show a regression curve and it's so normal for them\ldots they don't always realize that people don't understand some of the visuals for these models and what they really mean.}'' 

\smallbreak

\begin{itemize}
    
    \item[C1] { \textbf{Metrics and visuals provide little insight to SMEs without proper explanation, context, and rationale from data scientists.}
        Overall, data scientists more frequently brought up using metrics to illustrate and interpret the performance of a model than SMEs. Our usage of the `metric performance' code had the greatest difference of median tags across any code, with the median data scientist mentioning `metric performance' 11 times, while the median SME mentioned `metric performance' only 3.5 times. This finding supports previous literature that suggests data scientists, in practice, too often rely on metrics to communicate model performance\cite{shah2019making,seneviratne2020bridging, cresswell2013organizational}.
        \smallbreak
        
        SMEs reported uncertainty towards how error metrics relate to their end-goals for the model, and how metric performance should be interpreted: ``\textit{I've seen metric scores when working with data scientists. I've also observed others and myself wanting to know what was behind the score, how it was derived. I would say there's a great deal of skepticism. I think it's something that needs to be described or defined fairly quickly upfront\ldots where does it come from? What's the top score, what's the bottom score?}'' (PID11). When asked how data scientists could make metrics more interpretable, SMEs cited ``visuals'' as one possible method, so long as they are contextualized: ``\textit{Data scientists need to explain what the metrics mean. Visuals can be a good way to explain it. It's more that you can see the results on the graph and you're able to relate the results to the context that we're working in''} (PID12). However, some SMEs found visualizations used by data scientists to be confusing, typically due to lackluster explanations for what was being depicted in the chart.
    }
    
    \smallbreak
    
    \item[C2] { \textbf{Interactive presentations are valued by SMEs, but may displace focus from a data scientist's intended takeaways.}
        Although data scientists cited metric performance as their primary method for communicating a model's results, a few told us they had prepared demos for presentations in previous projects. These demos consisted of interactive components of the model that could be manipulated by SMEs (e.g., interactive cells in a computational notebook), or visualizations that depicted model performance. In general, SMEs highly valued these interactive and in-depth presentations on a model, 5 out of 6 SMEs told us they would prefer an interactive presentation over a static or short presentation. In some cases, SMEs did not have prior experience with interactive demos or presentations, however, every SME wanted the ability to ``play with the model,'' \textit{e.g.}, by seeing how the model's outputs are affected by particular inputs:
        ``\textit{During this presentation\ldots as I was observing what the data scientists were doing, I was thinking I'd really like to put in my own values to the model, rather than having them pre-defined by somebody. I would like to pop in my own things and see what it returns}'' (PID11).
        
        \smallbreak
        
        However, when asked whether model performance presentations should be interactive, one data scientist (PID1) told us that SMEs can focus on the wrong aspects of a demo or visualization: ``\textit{Sometimes I notice when we show a demo [of the model], which contains some graphs or some nice visualizations, SMEs can be drawn more to the visual aspects of the demo and not how the data was collected, or how the model actually performs\ldots They're more drawn towards the visual part, so some key assumptions are not asked\ldots Instead, there's this interest in like, can we color it differently? Can we change the labels?}'' This misplaced focus on the visual aspects of a presentation, rather than its depicted content, may be one cause for a lack of visualizations found in data science presentations\footnote{\url{https://hbr.org/2019/01/data-science-and-the-art-of-persuasion}}.
    }
\end{itemize}

\subsection{Understanding Model Strengths and Weaknesses}
The second theme focuses on major concerns raised by both data scientists and SMEs in understanding the caveats, edge cases, outliers, and limitations of a model. Many data scientists and SMEs told us that when a model's weaknesses (or limitations) are not fully communicated, it can delay or prevent model adoption -- ultimately disappointing the end-users of the model: \textit{``We need to bridge the gap between business expectations, what a machine learning model should be doing, and what realistically a data scientist can or cannot do.''} 

\smallbreak

    \begin{itemize}
    
        \item[C3] { \textbf{SMEs need comprehensive details for a model's weaknesses at all stages of development; data scientists need context for why a model is underperforming.}
        SMEs unanimously agreed it is essential to know how and when a model may not meet their needs, particularly at the beginning of a collaboration. SMEs generally referred to this information as the ``limitations'' or ``weaknesses'' for a model: ``\textit{Even at the beginning stages, it's still beneficial for our team to know what the model is all about. What are the strengths of this model, but also what are its weaknesses?\ldots We may re-align our business needs around those strengths and weaknesses. But if there's a business need which the model cannot fulfill, we need to know during a presentation. We can help figure out how to change the approach, or if we should look into other models}'' (PID8). Data scientists often remarked that their understanding of a model's weaknesses and limitations hinged on SMEs' ability to communicate nuanced, domain-specific context (\textit{e.g.}, outliers).
        
        \smallbreak
        
        Understanding the weaknesses of a model can also build confidence in SMEs when making decisions from the model's predictions. This is especially critical when the model is being used in a high-risk environment. PID9, who discussed the usability of a regression model that predicts water quality, was primarily concerned with when the model might fail: ``\textit{If we're using this model to tell people that water is potable, that means they're going to use it for washing, cooking, cleaning, drinking\ldots and if there's something wrong with it, there are consequences to that decision. You know, there's a human being at the end of it. That's why I'd want as much information about the model as possible. So if I'm going to make a decision about applying this in the real world, then at least I know exactly what its limitations are before making a decision.}'' 
        }
    
    \smallbreak
    
    \item[C4] { \textbf{Misunderstanding or misrepresenting the data used in training can result in downstream issues. }
        While both SMEs and data scientists expressed great interest in the technical details for the data (\emph{e.g.}, distributions, summary statistics), only data scientists commonly noted that it was important to know how the data was pre-processed. These pre-processing steps included transformations, splits, and imputation strategies. However, it was unclear whether these procedures are commonly communicated to SMEs -- despite the SMEs' domain knowledge for whether those steps seem reasonable to perform.
        SMEs were also concerned about the quality of the data (\textit{e.g.}, whether data was missing or there were outliers in the data). PID8 stated that data quality issues may at first glance look like legitimate outliers from a model: ``\textit{A high number of outliers show up incorrectly\ldots and when you start investigating you identify that it's because of data related or data quality issues.}'' In general, both data scientists and SMEs remarked that information about outliers, error, and representations of the data are critical to know when understanding a model's performance -- and without this understanding, downstream issues can occur that prevent or halt model adoption (\emph{e.g.}, the effects of data cascades\cite{sambasivan:2021:cascades}).
        
    }
    
    \smallbreak
    
    \item[C5] { \textbf{Mismatched criteria for model acceptance can lead to unmet expectations for a model's performance.}
        A common pain point for model collaboration, most often brought up by data scientists, was the lack of clarity in what made a model ``good'' or ``acceptable'' to SMEs. On one hand, data scientists felt they had satisfied most of (if not all) the criteria that they established with SMEs at the start of their collaboration: \textit{``SMEs don't necessarily understand the kind of undertaking you performed to improve a model. Perhaps for you it was an awesome result of a multi-month effort and you never thought you could do it. And then they say, isn't it still too much [error]?''} (PID4). 
        
        \smallbreak
        
        This concept of unmet or unrealistic model performance expectations is a common problem experienced by data scientists in machine learning collaborations\cite{passi2018trust}. However, SMEs expressed their own frustration and confusion when models failed to perform up to their standards when all other criteria seemed to be met: \textit{``We constantly need to revisit these kinds of discussions to understand the model better\ldots There is a perception to us that, OK, we've given data scientists everything they need. The variables, the data points. So why is the model not giving us the expected results?''} (PID8).
    }
      
    \end{itemize}

\subsection{Gaps in Language, Context, and Comfort}
Finally, we discuss mismatches in language as well as a lack of context and comfort experienced by SMEs in data science collaborations. A common sentiment shared by SMEs was feeling confused, overwhelmed, or anxious when discussing details for a model and its performance -- largely due to data scientists' inability to relate to SMEs: ``\textit{Something important for data scientists to understand is that they are potentially introducing a new concept to us or a new way of doing things, and sometimes they don't consider that as a reference point\ldots I've heard data scientists comment that some of us don't know what we're doing or that we don't understand data. So there can be a little bit of an aloofness to the role of the data scientist.}''  

    \begin{itemize}
    
    \item[C6] { \textbf{Overly technical concepts that are not properly communicated by data scientists may discourage SMEs from asking clarifying questions.}
        SMEs expressed that they had felt confused or overwhelmed by data science presentations during previous collaborations, largely due to unfamiliar language, metrics, data, and visualizations used by data scientists. 
        One SME noted that, as a result, she did not always feel confident asking questions during data science presentations: ``\textit{Sometimes these presentations just go over your head, and I think the end-user a good chunk of the time would be too embarrassed to say, I don't get what you're talking about}'' (PID11). In Section~\ref{sec:interviews:methodology:computation}, we examine language barriers that may be causing confusion between the two groups. 
        
        \smallbreak
        
        When we asked data scientists how they approach mismatches in the language spoken, many of them expressed comfort with explaining or defining concepts during presentations. One data scientist told us she includes slides in the `appendix' portion of presentations with definitions prepared: ``\textit{Definitely there are a few people in the audience who won't understand these mathematical terms, so we create slides that explain them in the appendix. If someone has a question on terms, we just take it to those slides}'' (PID5). Although back-up slides may help with providing definitions for terminology, it is unclear whether SMEs will always ask the necessary clarification questions to prompt those slides.
        
        \smallbreak
        
        In contrast, SMEs told us they prepare pre-read documents to help contextualize unfamiliar concepts to data scientists. These documents included definitions and examples for the SME's domain, their objectives, known data outliers, acronyms/terminology commonly used, and expectations for the model: ``\textit{What has worked well for me in the past is sending a document, whether it's an email or Word doc or PowerPoint, kind of outlining what we're seeing, what our expectations are, what our concerns are. Because we all live in our own worlds\ldots Data scientists are good at what they do, but maybe we [SMEs] aren't the best at conveying the full idea to someone who's not living and breathing in our space}'' (PID10).
    }
    
    \smallbreak
    
    \item[C7] { \textbf{A lack of illustrative examples disconnects the model from its proposed use case.}
        When discussing how data scientists can better connect a model to its proposed use case, SMEs most often requested practical, illustrative, and grounded examples of using the model to meet their objectives. 
        On one hand, SMEs seemed skeptical that a new model being proposed could in fact be better than their current practices, particularly when data scientists do not take the time to compare both approaches: ``\textit{If data scientists explained what our [current] method does and how it works, and why and how they've built their model\ldots if they told us the reason why their model could be better than ours, or how it could be more trustworthy, it would really help us understand that, OK, this new approach might actually be better}'' (PID8).
        
        \smallbreak 
        
        Further, without a descriptive or exemplifying translation for how a predictive model relates to the business's needs, SMEs can fail to see its practicality: ``\textit{Data science, predictive models,\ldots they're all kind of trendy at the moment. I would be conscientious about using a new model, making sure that the model that we would use aligns with the purpose and objective of what we're trying to accomplish}'' (PID9). When a model is being proposed, one SME suggested that stories can be crafted by data scientists to better illustrate its eventual use: ``\textit{It can be any story. It can be a weird story. It doesn't need to be real, it doesn't need to have happened yet. Anything that would help us imagine how your model could help us predict a real outcome based on the information in your story}'' (PID13).    
    }
    \end{itemize}

\definecolor{persianred}{rgb}{0.8, 0.2, 0.2}
\definecolor{oliveweb}{rgb}{0.42, 0.56, 0.14}

\section{Guidelines for Communicating Model Performance}
\label{sec:guidelines}

Based on the analysis of our interviews, we derive a set of guidelines that can begin to bridge common communication gaps when data scientists present a model's performance to subject matter experts.  For the sake of clarity, they are grouped broadly by the challenges they address (C1-C7).  We note that some guidelines may partially address many findings, and provide justification when that is the case. 

\smallbreak

The first three guidelines relate to the choice of \textbf{\textcolor{t10h_10_lblue}{presentation and visualization styles}} when communicating and presenting model performance to SMEs, with a focus on contextualizing visualizations and metric performance: ``\textit{Some people don't have experience with visualization outside of BBC infographics\footnote{\url{https://www.bbc.com/future/tags/infographic}}. 
I do realize it can be hard for me to remove my data scientist hat and put myself into the role of somebody who's not looking at a log plot every day.}'' 

\begin{itemize}[leftmargin=*,topsep=4pt, partopsep=0pt,itemsep=3pt,parsep=2pt,label={}]

\item
\hlgnine{G1}: {\textbf{Provide context for a model's performance by annotating plots with clear stories.} Each annotation serves to decode the intended message of the visualization, beyond the visualized data and legend. Providing a clearcut story in the presentation of a model gives the data scientist more control over the message being communicated, and makes that message more palatable to the SME (\textbf{C2, C6}).  Annotating visualizations that showcase individual data instances also provides the opportunity to highlight illustrative, concrete examples (\textbf{C7}).  By guiding the audience through sensible conclusions on a provided visualization, an SME could more quickly arrive at new conclusions with the same visualization, improving visualization literacy\cite{borner2019data} over time.
That said, building stories from model performance is a nontrivial task and an active area of research; we point to existing work for more in-depth guidance on audience-based AI/ML narratives\cite{wang2019designing, segel2010narrative}.   
}

\item
\hlgnine{G2}: {\textbf{For any chart that communicates a model's performance, provide a range of comparisons.} Providing a comparison helps focus the message of a presentation (\textbf{C2}), in particular on how the model might improve their current practices (\textbf{C3, C7}).  It also helps define what qualifies as ``good" and ``bad" performance (\textbf{C5}).  In our study, SMEs found that assessing the results of a predictive model's performance is easier if it is compared against their current practices, in addition to an interpretable naive baseline model. If possible, an oracle, perfect model (or otherwise ground truth) can also be used for comparison. 
}

\item
\hlgnine{G3}: {\textbf{Visually explain the significance of aggregated metrics.} Global metrics such as explained variance or mean absolute error can seem abstract and removed from the use case (\textbf{C1}).  By sharing a visual medium, data scientists and SMEs can use it as a common language to discuss the significance of the metrics of a model.  Showing metrics in visual context can help ground them; for example, visualizing the enveloping ellipse in a correlation scatterplot can give a proxy for the correlation between predicted and actual values.  Whenever possible, showing how individual instances relate to the aggregated metric can help connect an SME's understanding of \textit{items} of data with the metric calculated on the full \textit{set} of data, which partially addresses the need for illustrative examples (\textbf{C7}).
}

\end{itemize}

The second three guidelines address concerns by both data scientists and SMEs in understanding the \textbf{\textcolor{persianred}{caveats, edge cases, outliers, and limitations}} of the model: ``\textit{If data scientists said, `when you run these models, here is the area where we think you're going to have the most problems, or the most risk. And here's the explanation for why we think that's happening.’\ldots I think upfront and transparent communication about why we should expect those issues is a very big way for us to build trust and confidence in the model.}''

\begin{itemize}[leftmargin=*,topsep=4pt, partopsep=0pt,itemsep=3pt,parsep=2pt,label={}]

\item
\hlgfour{G4}: {\textbf{Point to outliers in the model's performance and data space with known or plausible explanations.} SMEs have deep knowledge of the data being used to train the model and make predictions, so it is a missed opportunity if outliers are not discussed between SMEs and data scientists (\textbf{C3, C4}). The source of outliers and anomalies is often dependent on the scenario, therefore, data scientists should point SMEs to known or potential outliers, and include at least reasonable speculations behind their anomalous behavior. Use visualizations that illustrate the biggest errors in model performance, and show the distribution of features to point to potential outliers.}

\item
\hlgfour{G5}: {\textbf{Explicitly describe the data used in training and testing a model, any pre-processing or transformations performed, and provide examples.} The distribution, weighting, correlation, and availability of the data used in the modeling process were notable concerns from both SMEs and data scientists (\textbf{C4}). Many data scientists agreed that SMEs provide essential context for the data domain, ultimately leading to improvements in model performance and transparent communication. 
In addition to explicit communication about the data used in modeling, we point to HCI solutions that can prevent the downstream effects of data related issues\cite{sambasivan:2021:cascades}. }

\item
\hlgfour{G6}: {\textbf{Communicate the limitations, weaknesses, and blind spots of the model, especially when suggesting it can replace the existing solution(s).} 
SMEs want transparent information regarding the limitations of a model (\textbf{C3}) with qualitative and quantitative assessments of those errors or weaknesses against their current practices (\textbf{C7}). Both SMEs and data scientists noted that they were able to help each other improve a model once the limitation or poor performance of the model was fully understood -- \emph{e.g.}, a model overfitting, or a linear model performing poorly on outliers -- ultimately leading to higher model acceptance.  By understanding the reasoning behind weak performance (data quality, outliers, weak signal), SMEs can better understand the criteria for assessing a model's performance (\textbf{C5}).}
\end{itemize}

The last three guidelines address a need for \textbf{\textcolor{oliveweb}{context and comfort}} identified by SMEs: ``\textit{You have to make the end-user feel comfortable both in the data scientist's language, and also that if they don't understand something they can ask, what is this?}'' 

\begin{itemize}[leftmargin=*,topsep=4pt, partopsep=0pt,itemsep=3pt,parsep=2pt,label={}]
\setcounter{enumi}{6}

\item
\hlgeight{G7}: {\textbf{When articulating results, start slow and offer to speed up; use visualization as a medium when appropriate.} Lack of comfort or understanding should be anticipated in a presentation of data science results, as SMEs have distinct backgrounds (\textbf{C6}). Data scientists need to be cognizant of this imbalance, particularly at the beginning of their presentations.  SMEs suggested that data scientists could spend more time highlighting aspects of their presentation that could be considered ``obvious,'' in order to establish a common baseline for the language spoken and understood. Using interpretable visualizations (\hlgVis{G1}) to drive the discussion can make it easier to catch common misunderstandings and ground the discussion. }

\item
\hlgeight{G8}: {\textbf{Provide a pre-read document or appendix with explanations and illustrations of any models, metrics, or unfamiliar vocabulary.} Since many data science concepts may be new to the audience (\textbf{C6}), it is beneficial to provide an option to read through both the definitions and the results, ideally offline before and after a presentation at the SME's own pace. 
Frisch et al. present details for successful pre-reads\cite{frisch2020takes}, and Crisan et al. describe how visualization can aid in data science documentation\cite{crisan2020passing}. If a pre-read is not possible, an appendix slide with definitions should be included and deliberately referenced in the presentation. 
}

\item
\hlgeight{G9}: {\textbf{Tie in use cases for the model by illustrating real-life, objective-driven examples.}} 
SMEs need to understand how a model's performance relates to their end-goals before recommending its use (\textbf{C7}). Illustrative use cases can include individualized examples that showcase how the prediction of the model is affected by particular inputs (\textbf{C2}), as well as examples of the model's predictions being used in real-world applications.  Similar to story-telling visualizations as in \hlgnine{G1}, illustrative examples can provide an entry point to engage the audience and provide comfort (\textbf{C6}).

\end{itemize}

\urlstyle{tt}

\begin{table*}[ht!]
    \centering
    \centering
    \caption{
    An illustration of how our guidelines (Section~\ref{sec:guidelines}) can be applied in practice to a data science presentation.
    Each row represents a presentation element (\textbf{D1}-\textbf{D4} and \textbf{V1}-\textbf{V4}) that was shown to SMEs during a PowerPoint presentation of two models' performance (details in Section~\ref{sec:demonstration}). The left column shows the guidelines being demonstrated, the middle column describes how the guidelines can be implemented, and the right column summarizes SMEs' open-ended feedback during our follow-up study, with quotes, on the effectiveness of the guidelines and visualizations.
    }
    \vspace{-10pt}
    \scalebox{0.86}{
    \renewcommand\theadalign{tl}
\renewcommand\theadgape{\Gape[4pt]}
\renewcommand\cellgape{\Gape[4pt]}

\begin{tabular}[t]{c p{0.25\linewidth} p{0.45\linewidth} p{0.3\linewidth} }
	\toprule
	{} &
    \textbf{Guidelines} & 
    \textbf{How the guidelines can be implemented} & 
    \textbf{SMEs' feedback on effectiveness}\\
    
    \midrule
    
    {\textbf{D1}} &
    (\hlgeight{G7}) Ask for clarification & 
    Slide instructing participants to ask clarifying questions about data, charts, or any language / vocabulary. & 
    Statement made the presentation feel like ``\textit{an exchange, instead of an exam.}''\\
    
    \addlinespace
    \hline
    \addlinespace
    
    {\textbf{D2}} &
    (\hlgfour{G5}) Data descriptions & 
    Table detailing all data attributes used in training, including units, a short description, and an example for each. & 
    Quickly familiarized SMEs with the data, ``\textit{otherwise they're just numbers.}''  \\
    
    \addlinespace
    \hline
    \addlinespace
    
    {\textbf{D3}} &
    (\hlgeight{G8}) Terminology definitions & 
    Comparative explanation for how different regression models make predictions. Definitions for common metrics (e.g., R$^2$, MSE) were included, with descriptions of high versus low error. &  
    Gave models and metrics context: ``\textit{Without descriptions, those [models] could be anything.}''\\
    
    \addlinespace
    \hline
    \addlinespace
    
    {\textbf{D4}} &
    (\hlgfour{G6}) Show model weaknesses and (\hlgeight{G9}) illustrate how these instances impact real-life objectives.
    &
    Specific examples -- tied to the prediction task and objective for the model -- that illustrate one model performing worse than another, with explanations and context for why.  
    & 
    Objective-driven examples were cited as the most helpful for decision-making by SMEs.\\
    
    \addlinespace
    \hline
    \addlinespace
    
    {\textbf{V1}} &
    (\hlgnine{G3}) Visually explain multiple metrics and (\hlgLimit{G4}, \hlgfour{G6}) see where they perform strongly or poorly.  (\hlgVis{G2}) Compare performance to a baseline and (\hlgnine{G1}) see annotated examples.
    &
        {
        \begin{minipage}[t]{.45\textwidth}
            Heatmap showing the apportionment of error across subgroups of the data, split by metric. Reveals potential bias in the data.\\
        \end{minipage}%
    }
    {
    \begin{minipage}[t]{0.22\textwidth}
        \centering
        \vspace{-2pt}
        \textbf{Baseline model} \\[3pt]
        \includegraphics[width=.9\linewidth]{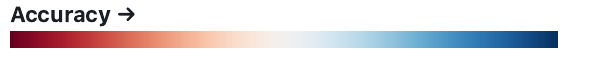}\\
        \includegraphics[width=.9\linewidth, trim={0 20pt 0 0}, clip]{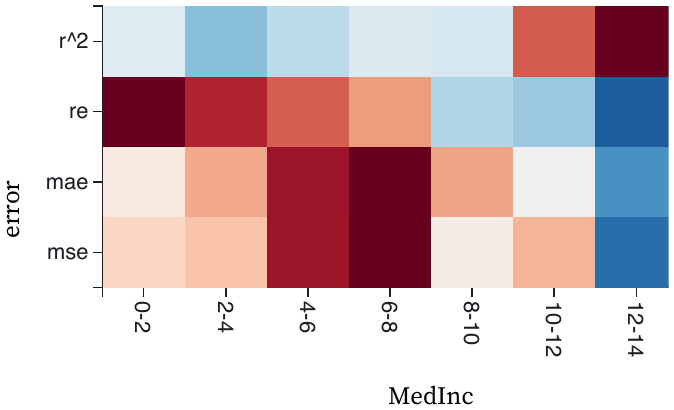}\\
        \includegraphics[width=\linewidth, trim={0 11pt 0 0}, clip]{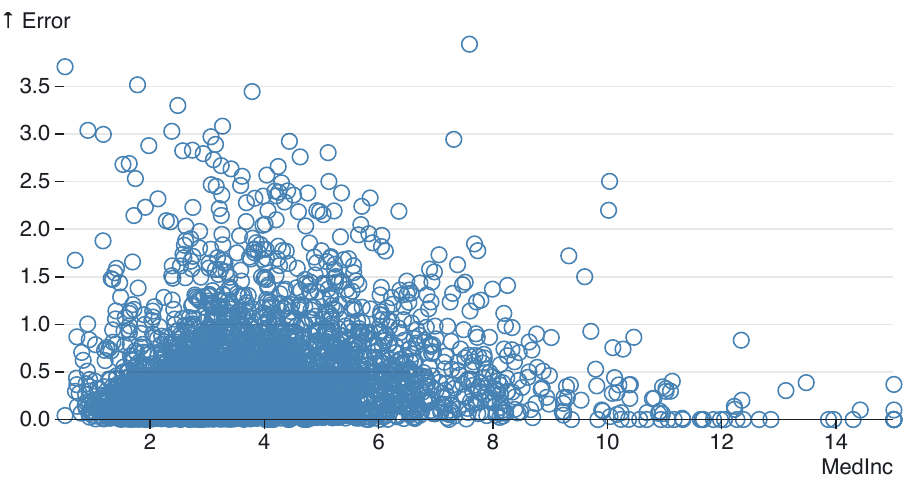}
        \end{minipage}%
    \hfill
    \begin{minipage}[t]{0.22\textwidth}
        \centering
        \vspace{-2pt}
        \textbf{New model} \\[3pt]
        \includegraphics[width=.9\linewidth]{figures/teaser/cali_xgb_heatmap_legend.pdf}\\
        \includegraphics[width=.9\linewidth, trim={0 20pt 0 0}, clip]{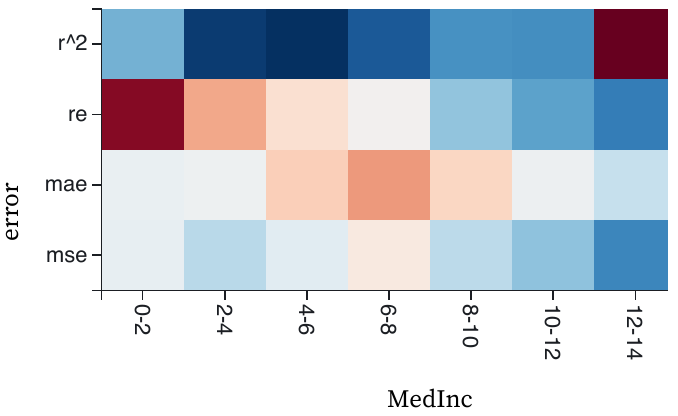}\\
        \includegraphics[width=.9\linewidth, trim={0 10pt 0 0}, clip]{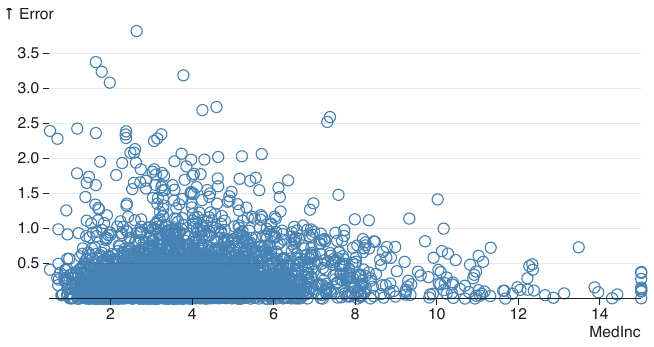}
    \end{minipage}%
    }
    &
    { 
    Easy to interpret the strengths and weaknesses of the models: ``\textit{The visual [heatmap] made it easier to draw conclusions, the numbers alone were not as easy to translate what the model means, or how you could apply it if you were really using it [for the task].}'' }\\
    \addlinespace
    \hline
    \addlinespace
    
    {\textbf{V2}} &
    (\hlgnine{G3})Visually explain R$^2$ metric and (\hlgfour{G4}) discover outliers in prediction error.  (\hlgnine{G2}) Compare performance to a baseline and (\hlgnine{G1}) see annotated examples.
    
    & 
    {
        \begin{minipage}[t]{.45\textwidth}
            Correlation scatterplot between predicted \& actuals to identify outliers. A containing ellipse provides a visual for the R$^2$ metric.\\
        \end{minipage}%
    }
    {
    \begin{minipage}[t]{0.22\textwidth}
        \centering
        \vspace{-2pt}
        \textbf{Baseline model}\\[3pt]
        \includegraphics[width=.9\textwidth]{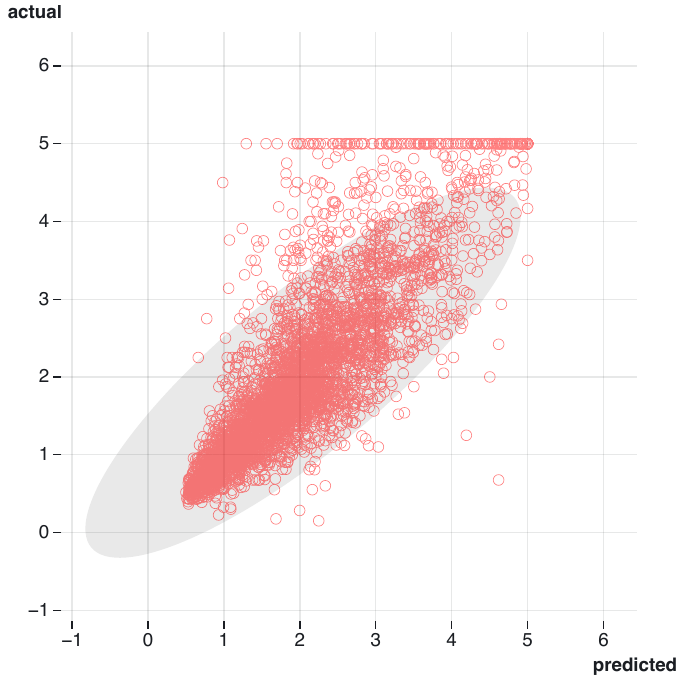}
        \end{minipage}%
    \hfill
    \begin{minipage}[t]{0.22\textwidth}
        \centering
        \vspace{-2pt}
        \textbf{New model}\\[3pt]
        \includegraphics[width=.9\textwidth]{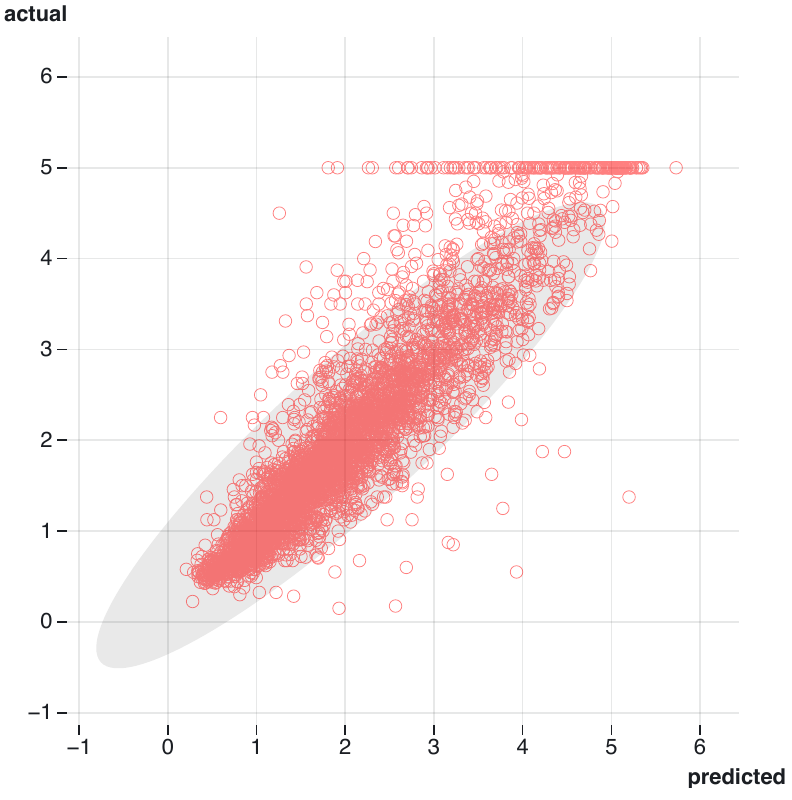}
    \end{minipage}%
    }
    &
    Useful for illuminating R$^2$. The bounding ellipse helped discern the difference in precision for the two models, as well as identifying outliers and potential risks: ``\textit{I found it particularly easy with the ellipse in the middle to gauge where each model was better. I can see visually where the difference was, the outliers.}''\\
    \addlinespace
    \hline
    \addlinespace
    
    {\textbf{V3}} &
    (\hlgfour{G4}) Discover outliers in prediction error and (\hlgLimit{G5}) view the difference in distributions.  (\hlgnine{G2}) Compare performance to a baseline and (\hlgnine{G1}) see annotated examples.
    
    & 
    {
        \begin{minipage}[t]{.45\textwidth}
            Paired histograms comparing the distribution of predicted values from the model versus the actual values from the raw data.\\
        \end{minipage}%
    }
    {
    \begin{minipage}[t]{0.22\textwidth}
        \centering
        \vspace{-2pt}
        \textbf{Baseline model}\\[3pt]
        \includegraphics[width=.9\linewidth, trim={0 15pt 0 0}, clip]{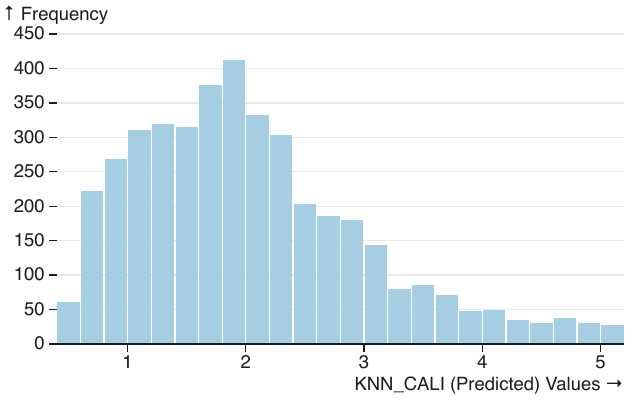}\\
        \textbf{Original Data}\\[3pt]
        \includegraphics[width=.9\linewidth, trim={0 15pt 0 0}, clip]{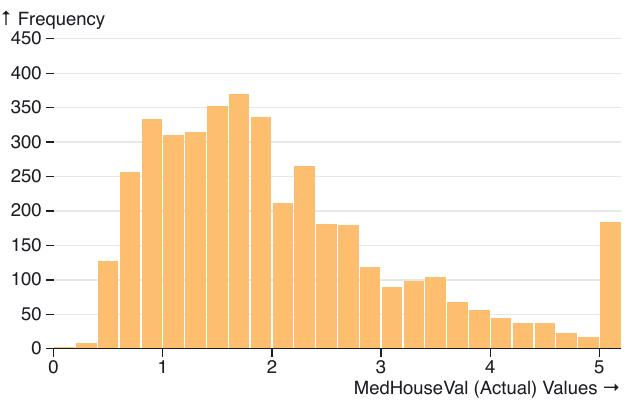}
        \end{minipage}%
    \hfill
    \begin{minipage}[t]{0.22\textwidth}
        \centering
        \vspace{-2pt}
        \textbf{New model}\\[3pt]
        \includegraphics[width=.9\linewidth, trim={0 15pt 0 0}, clip]{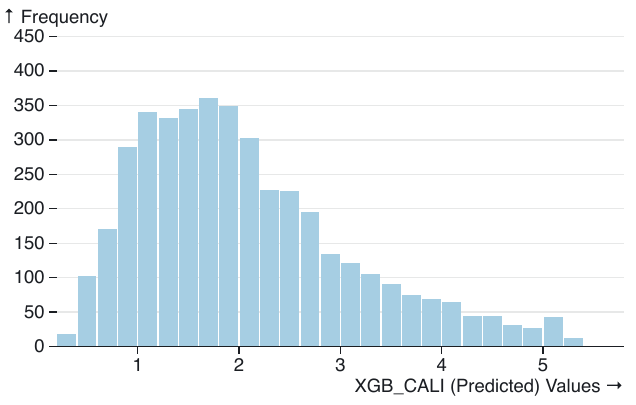}\\
        \textbf{Original Data}\\[3pt]
        \includegraphics[width=.9\linewidth, trim={0 15pt 0 0}, clip]{figures/teaser/cali_xgb_histogram_act_newcolors.pdf}
    \end{minipage}%
    }
    &
    Simplest visualization to grasp, helping SMEs see what types of errors the two models had compared to ground truth: ``\textit{Easy to compare the two side by side, the overall shape gives you an idea of the errors that can come up. It's just these two bars at the end each time.}'' \\
    \addlinespace
    \hline 
    \addlinespace

    {\textbf{V4}} &
    (\hlgnine{G3})Visually explain MAE metric and (\hlgLimit{G6}) observe where error is apportioned.  (\hlgnine{G2}) Compare performance to a baseline and (\hlgnine{G1}) see annotated examples.
    & 
    {
        \begin{minipage}[t]{.45\textwidth}
            Absolute error plots to contextualize the MAE metric (in red). Shows the density of data with better or worse average error.\\
        \end{minipage}%
    }
    {
    \begin{minipage}[t]{0.22\textwidth}
        \centering
        \vspace{-2pt}
        \textbf{Baseline model}\\
        \includegraphics[width=.9\textwidth, trim={5pt 45pt 0 0}, clip]{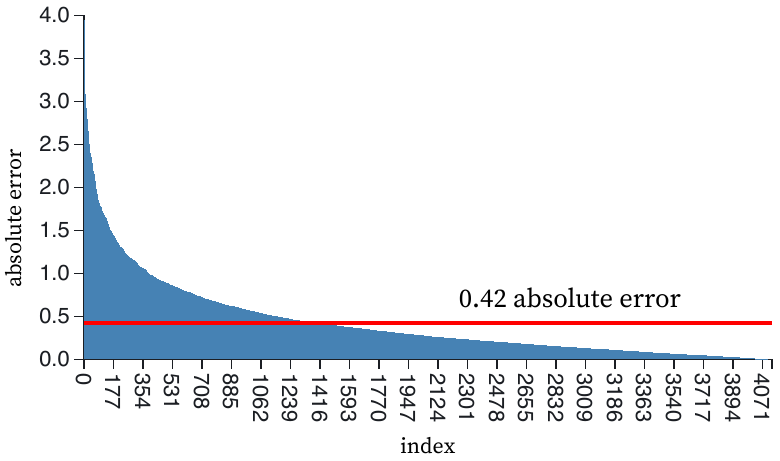}
        \end{minipage}%
    \hfill
    \begin{minipage}[t]{0.22\textwidth}
        \centering
        \vspace{-2pt}
        \textbf{New model}\\
        \includegraphics[width=.9\textwidth,  trim={5pt 45pt 0 0}, clip]{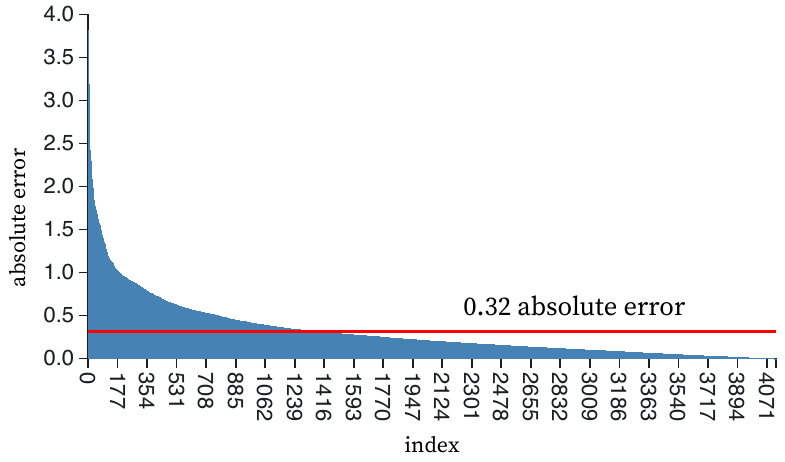}
    \end{minipage}%
    } \vspace{1pt}
    &
    { 
    Models were not visually distinguishable, making it unclear what finding was being communicated in the chart: ``\textit{Probably wasn't as useful in this exercise. But there are probably circumstances where it might be helpful as a more differentiating figure.}''}\\
    
    \addlinespace
    
    \bottomrule

\end{tabular}

}
    \label{tab:validation_results_guideline}
\end{table*}

\section{Demonstration of the Guidelines}
\label{sec:demonstration}

We demonstrate the use of our guidelines in a data science presentation with SMEs. For this demonstration, we developed two slide-based presentations on distinct regression modeling scenarios: the first scenario \textit{without} our guidelines, and the second scenario \textit{with} our guidelines. We conducted a second round of interviews with the SMEs from our first interview study (Section~\ref{sec:interviews}), treating the first presentation as a point of comparison to elicit feedback on our guidelines.

\subsection{Follow-up Interviews}

From our first interview study, a common modeling scenario experienced by SMEs involved meeting with a data scientist to assess whether a newly developed model was ready for deployment or needed to be more performant. Error metrics were typically used by data scientists to depict the model's performance, but metrics alone failed to capture key aspects of the model, such as its risk-benefit ratio, or the practical value of its outputs. Consequently, SMEs struggled to determine how well the model would perform in real-world situations, or how it could produce meaningful insights.

Therefore, the objectives of our follow-up study were to understand whether our proposed guidelines could help inform and contextualize the evaluation of a model during a data science presentation, while also demonstrating how data scientists can incorporate our guidelines into their own presentations to better communicate the accuracy, strengths, and weaknesses of a model's performance to SMEs.

The slides we presented to SMEs in this follow-up study are available as supplemental material, including all visualizations and data tables seen by participants, and can be accessed at \url{https://github.com/TuftsVALT/ModelComm}. Below, we describe the protocol for our demonstration of the guidelines.

\begin{itemize}[leftmargin=0pt,topsep=4pt, partopsep=0pt,itemsep=2pt,parsep=2pt]
    \item[] {\textbf{Participants:}
    To understand whether our guidelines address the challenges identified from our first interview study (Section~\ref{sec:interviews}), the six SMEs who participated were contacted for a follow-up study approximately six months later, with four agreeing to participate. Interviews were held virtually in a one-on-one one-hour session.  Each session walked through a set of slides describing two distinct modeling scenarios.}
    
    \item[] {\textbf{Scenarios:} Scenarios 1 and 2 described different datasets with similar modeling goals. 
    In both scenarios, participants listened to a presentation of a new model being proposed to replace a model currently in use. For the first scenario, designed as a control setting, a baseline set of information was given to communicate model performance based on common techniques described by the data scientists in our first interview study, as well as our review of the related literature.  In the second scenario, the experimental setting, the presentation of model results was augmented according to our guidelines, including visualizations. Because this follow-up study consisted of only four participants, we did not switch the order of the scenarios presented during the interviews. To balance the ordering, we used a more complex dataset and prediction task in scenario 2.}
    
    \item[]{\textbf{Datasets:} 
    For scenario 1, we used the Auto MPG dataset\cite{donoho1982autompg} (6 features) to train a regression model to predict a vehicle's miles per gallon (MPG). For scenario 2, we used the California housing dataset\cite{pace1997calihousing} (9 features) to train a regression model to predict a house's cost. All features in the original datasets were used in training our models.} 
    
    \item[]{\textbf{Models used:} SMEs were told that a KNN\cite{fix1989knn} (``baseline'') model was already in use, and were asked to assess whether an XGB\cite{chen2016xgboost} (``new'') model should be used instead. In both scenarios, the XGB model had improved performance.}
    
    \item[]{\textbf{Prediction tasks:} 
    Both tasks asked SMEs to choose between the KNN or XGB model. In scenario 1, SMEs were told they would use their chosen model to predict the MPG of a vehicle based on its specifications. In scenario 2, SMEs were told they would use their chosen model to predict the value of a property for a future investment. }

\end{itemize}

\subsection{Presentations}

\noindent 
\textbf{Scenario 1 (no guidelines):} 
At the start of both scenario presentations, we included a short description of the dataset, the model inputs, and the SME's prediction task. An example of the input data was provided as well as a table of the first 10 prediction results for the KNN (baseline) and XGB (new) models. Following the most standard practices described by data scientists and SMEs in our first interview study, we did not include any visualizations for the first scenario's presentation. Instead, a table of common regression model metrics (R$^2$, MSE, MAE) were displayed to illustrate the performance of both models. 

\noindent 
\textbf{Scenario 2 (with guidelines):} 
Before starting our second scenario, we explained to SMEs that supplementary information would be included in the next set of slides.  This additional information was designed to reflect the guidelines described in Section~\ref{sec:guidelines}; the description of the mapping is shown in Table~\ref{tab:validation_results_guideline}.  In total, there were four differences (\textbf{D1}-\textbf{D4}) incorporated into the presentation of our second scenario. This additional information was embedded and explicitly highlighted in our slides, and can be found in our provided supplemental. We briefly describe each below: 

\begin{itemize}[topsep=2pt, partopsep=2pt,itemsep=2pt,parsep=2pt]
    \item[\textbf{D1}] { \textbf{Open to questions:} An initial slide encouraging SMEs to stop us at any time to ask clarifying questions. (\hlgContext{G7})}
    
    \item[\textbf{D2}] { \textbf{Detailing the data:} A table that descriptively defined all features used in training and testing the models. (\hlgLimit{G5})}
    
    \item[\textbf{D3}] { \textbf{Defining all terminology:} High-level descriptions of the modeling algorithms (KNN and XGB) and how they differ in deriving their predictions. Definitions for R$^2$, MSE, and MAE were supplied. These definitions included how a lower or higher value for each metric translate to relative performance. (\hlgContext{G8})}
    
    \item[\textbf{D4}] { \textbf{Highlighting model weaknesses:} Several examples of XGB performing worse than KNN, on both instance-levels and visualized regions of data. Each example included a short description to explain why we believed XGB was predicting worse than KNN for these specific examples, and how this performance would relate to real-world objectives (\hlgLimit{G6}, \hlgContext{G9}).}
\end{itemize}

Lastly, we used four different visualizations (\textbf{V1}-\textbf{V4}) to illustrate differences in performance for the two models. Our chosen visualizations were based on the guidelines and deliberated between all authors; however, we note that a broad design space of visual encodings would pertain to our guidelines. The four chosen for this demonstration were picked to illuminate several key aspects of the guidelines: showing outliers, enabling comparisons, and providing both global and instance-level views into metrics.  We considered charts that are recognizable and interpretable to SMEs (\emph{e.g.}, heatmaps, scatterplots, histograms) that can also address these model communication needs.

\urlstyle{same}

Every visualization for the XGB (new) model included a comparison visualization to the KNN (baseline) model (\hlgVis{G2}). For each visualization, one or more annotations were included (\hlgVis{G1}) to point to specific outliers (\hlgLimit{G4}), differences (\hlgContext{G8}), and conclusions (\hlgContext{G9}) between the two models. We read the annotations to SMEs during the presentation and asked if they had any follow-up questions. Full versions of the visualizations can be seen in Table~\ref{tab:validation_results_guideline}, all source code is provided in an Observable notebook\footnote{\url{https://observablehq.com/@ashleysuh/model-comm-vis}}, and the annotations are included in our supplemental. The visualizations used in our follow-up are described as follows: \\

\noindent 
\textbf{(V1) Heatmap by error category and feature distribution: } 
\begin{wrapfigure}{l}{0.1\textwidth}
    \centering
    \includegraphics[width=0.1\textwidth]{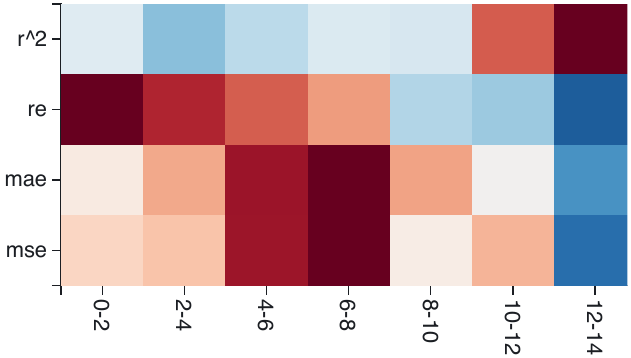}
    \includegraphics[width=0.1\textwidth]{figures/teaser/cali_knn_fdp.pdf}
\end{wrapfigure}

\noindent 
A heatmap that illustrates aggregate errors (y-axis) across a single feature (x-axis), using a distinct color scale for each row (\emph{i.e.} error metric). The heatmap can help SMEs see how different aggregate measures (\hlgVis{G3}) differ across groups (\hlgLimit{G4}) and determine if a model's performance has weaknesses along the selected feature (\hlgLimit{G6}). This is especially important when looking at how a model performs across regions of data, ensuring that a model with a slight increase in overall accuracy does not sacrifice performance within a specific subgroup of data. Attached to the heatmap is a one-dimensional scatterplot showing the distribution of the feature used to create the grouping, to allow for viewing individual instances when interpreting aggregate measures. \\

\noindent 
\textbf{(V2) Correlation scatterplot with confidence ellipse:}  

\begin{wrapfigure}{l}{0.1\textwidth}
    \centering
    \includegraphics[width=0.1\textwidth]{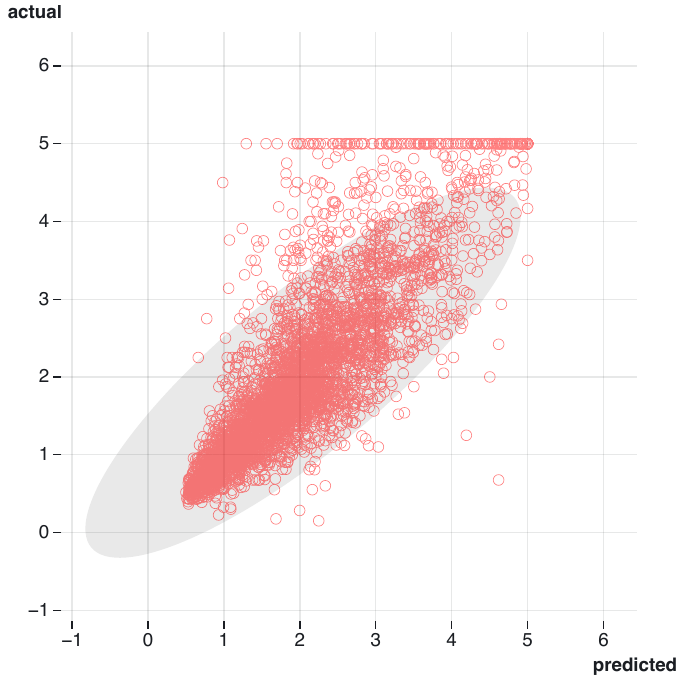}
\end{wrapfigure}

\noindent 
A scatterplot with predicted values from the model plotted on the x-axis, and actual values plotted on the y-axis. Behind the plotted points is a 95\% confidence interval (or covariance ellipse\cite{covarianceEllipse}) that serves as a graphical proxy\cite{friendly2013elliptical} for the regression metric R$^2$ (\hlgVis{G3}).  Simultaneously displaying the aggregate metric R$^2$ along with individual instances in the scatterplot provides a visual interpretation for the R$^2$ metric.  In addition, it makes it easy to see outliers in predictions and labels (\hlgLimit{G4}). \\

\noindent
\textbf{(V3) Histogram for predicted values versus actual values:}

\begin{wrapfigure}{l}{0.1\textwidth}
    \centering
    \includegraphics[width=0.1\textwidth, height=.96cm]{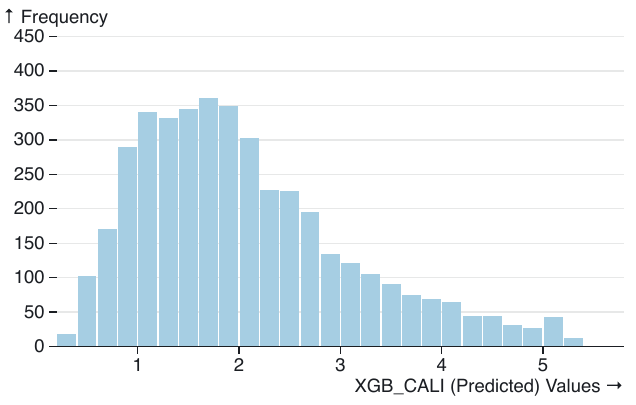}
    \includegraphics[width=0.1\textwidth, height=.96cm,]{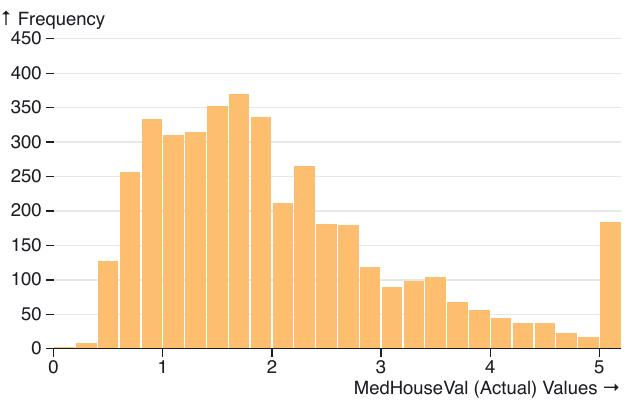}
\end{wrapfigure}

\noindent 
Two one-dimensional histograms, each showing the distributions of a model's prediction and actual values, respectively (\hlgLimit{G5}).  By highlighting the differences in shape of predicted versus actuals, this visualization makes it easy to see if a model might have biases (\textit{e.g.}, missing the high-range outliers found in the ground truth, or missing one of the modes of the distribution of actual values) (\hlgLimit{G4}). \\

\noindent 
\textbf{(V4) Bar chart with global error:}  

\begin{wrapfigure}{l}{0.1\textwidth}
    \centering
    \includegraphics[width=0.1\textwidth]{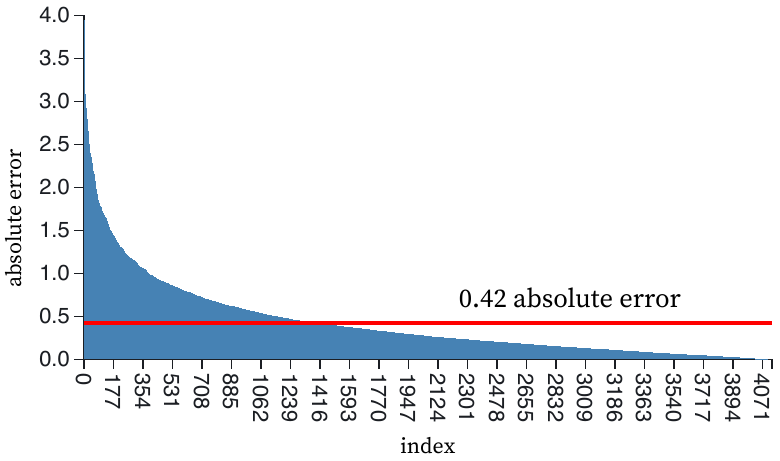}
\end{wrapfigure}

\noindent 
A bar chart where each bar represents the absolute error of a single instance; bars are sorted by descending error.  This chart shows the \textit{shape} of error by the models, and conveys how it apportions error, whether by apportioning it equally among all instances, or confining it to few (\hlgLimit{G6}).   A red horizontal line is added to represent the mean absolute error, giving more context to the per instance error for the MAE and MSE metrics. (\hlgVis{G3})

\smallbreak 

\textbf{V1} and \textbf{V3} were used as part of the presentation of examples where the model performed poorly (\textbf{D4}). Any time XGB performed worse than KNN in our examples, we included a short description to explain why we believed XGB was predicting worse than KNN (\hlgLimit{G4}).

\medbreak 

\noindent \textbf{Feedback questions:} 
    At the end of our demonstration, we asked SMEs to evaluate the effectiveness for each of the presentation differences (\textbf{D1}-\textbf{D4}) and visualizations used (\textbf{V1}-\textbf{V5}). SMEs rated their agreement to the following statement:
    \begin{itemize}[leftmargin=12pt,topsep=2pt, partopsep=0pt,itemsep=2pt,parsep=2pt]
        \item[] { ``\textit{I found this difference / visualization to be helpful in interpreting the model’s performance.}''}
    \end{itemize}
    \noindent 
    Agreement was made on a 7-point Likert scale of 1 (Strongly Disagree) to 7 (Strongly Agree). Finally, SMEs provided open-ended feedback on the helpfulness of the guidelines.

\subsection{SME Feedback}

\subsubsection{Additional model information (D1-D4)}
In general, all supplemental information for the models (\textbf{D1}-\textbf{D4}) were considered helpful by SMEs, which is not surprising on its own: contextual information is generally helpful during model communication, as evidenced by challenges C6 and C7 from Section~\ref{sec:findings}, and previous work on improving model adoption by SMEs\cite{shah2019making}. 

Ranked by average helpfulness score, \textbf{D4} [\textit{Highlighting model weaknesses}] (6.75) was most helpful, reflecting our finding that understanding the strengths and weaknesses of a model (\hlgLimit{G6}), as well as seeing objective-driven examples of its predictions (\hlgContext{G9}), impacts and informs decision-making.  It was followed by \textbf{D2} [\textit{Detailing the data}] (6.5) and \textbf{D3} [\textit{Defining all terminology}] (6.25), which provide additional context on the raw data, the models, and the metrics used in a presentation (\hlgLimit{G5}).  The lowest rated difference \textbf{D1} [\textit{Open to questions}] (5.5) was split, with two SMEs rating it as not helpful or unhelpful, and two SMEs rating it as strongly helpful. Below, we distill takeaways from the use of \textbf{D1}-\textbf{D4}.

\begin{itemize}[leftmargin=0pt, topsep=1pt, partopsep=2pt,itemsep=2pt,parsep=2pt]
    \item[] { \textbf{Highlighting the model's weaknesses informs decision-making and boosts credibility:}    
    SMEs strongly preferred examples of the model performing poorly. PID9 explained that they helped her ``\textit{come to the conclusion about whether there's a real performance difference between the two models},'' reinforcing our previous finding that SMEs need transparency from data scientists on a model's weaknesses. Similar to challenge C3 identified in Section~\ref{sec:findings}, PID8 told us that seeing the model's limitations would help him decide whether the model could be improved, or would be a no-go for future tasks: ``\textit{Explanations around why KNN shows such worse results than XGB\ldots That helps us try to see if it's a valid reason, or if it's a straight no-go and we cannot consider the model.}''}
    
    \item[] { \textbf{Describing the data used in training ``beyond numbers'' level-sets the presentation:} 
    SMEs found that detailed descriptions and examples of the modeling data provided essential context beyond commonplace summary statistics: ``\textit{Always show this [data descriptors] before the numbers. Otherwise they're just numbers}'' (PID11). PID8 had similar feedback, remarking that ``\textit{these descriptions will help everyone [SMEs] be on the same page as the data scientists},'' thus grounding the model and its performance to the real world.}
    
    \item[] { \textbf{Defining data science terminology, model differences, and error metrics establishes a common framework:} 
    Overall, SMEs found the added definitions and descriptions of the models used (KNN and XGB) to be helpful in interpreting how predictions are made. PID9 told us, ``\textit{without your descriptions, KNN versus XGB could be anything. General information is helpful.}'' 
    Both PID9 and PID11 suggested that walking through examples with ``\textit{named attributes}'' in a ``\textit{nontechnical context}'' could be even more helpful (\hlgeight{G8}). 
    
    Although some SMEs were already familiar with the metrics used in our presentation, all SMEs found the added description for what `low' and `high' error means to be helpful in contextualizing model performance: ``\textit{I already know about these [R$^2$, MSE, etc.], what's good and bad. But of course knowing what high versus low error means is helpful}'' (PID11). This finding reinforces challenge C1 and guidelines \hlgVis{G2} and \hlgVis{G3}: ``\textit{I don't use R$^2$ or MSE or any of these on a day to day basis. In our world, we use p-values\ldots It's difficult at times to know in which scenarios lower is better or higher is better}'' (PID9).}
    
    \item[] { \textbf{Encouraging SMEs to ask for clarification helps the presentation feel conversational:}
    The overall sentiment towards our slide that encouraged clarification questions was split. PID11 told us, ``\textit{I think I'd already be proactive if there was something I didn't understand.}'' On the other hand, PID9 felt the addition made our presentation more conversational: ``\textit{The perception it creates when you have this statement is that it's more of an exchange. Without it, it feels like more of an exam.}'' PID8 related it to having a two-way, as opposed to one-way, presentation: ``\textit{It does give you the liberty that nothing is written in stone, it's not something that's one way.  It's kind of a two-way as you go through [the presentation]\ldots  You're not just saying, this is what we are doing.}'' We believe the additional step by data scientists to remind SMEs that they can always pause the presentation to ask questions will help encourage those who may not otherwise (\hlgeight{G7}).}
\end{itemize}

\subsubsection{Model performance visualizations (V1-V4)}

Visualizations \textbf{V1} [\textit{Heatmap}] (7.0), \textbf{V2} [\textit{Scatterplot}] (6.5), and \textbf{V3} [\textit{Histogram}] (6.0) were all considered helpful, while \textbf{V4} [\textit{Bar chart}] (2.75) was not considered helpful. Overall, we find that \textbf{visual comparisons that clearly tell a story are preferred to simple text and metrics.} All SMEs pointed out that with the supplied visualizations, they had ``\textit{quick}" or ``\textit{instant}" insights into model performance, whereas text or numbers had to be studied.  Below, we discuss the qualitative feedback (and overall Likert ratings) received from SMEs on the visualizations shown in our presentation.

\smallbreak 

The heatmap \textbf{V1} was found to be ``strongly helpful'' in interpreting the models' performance for all participants, partly because of its easy interpretation: ``\textit{Easily interpretable since the colors pop out, colors are easier to draw conclusions from and find outliers}'' (PID11); ``\textit{I can immediately see which model is better and which is worse}'' (PID8), and partly because it helped communicate the strengths and weaknesses of the models: ``\textit{This one gives you even more detail as to how the models are behaving, where they're having the most errors}'' (PID9).  

The scatterplot \textbf{V2} was found to be useful for understanding the metric R$^2$, as PID13 noted that the ellipse helped them see the difference in precision for the two models: ``\textit{With the ellipse in the middle, it was easy to gauge where each model was better},'' but it was also helpful for identifying outliers and alerting risks by PID8 and PID9: ``\textit{I can see visually where the difference was, the outliers -- the values falling outside of the ellipse on the left side.}'' 

The histogram \textbf{V3} was considered the easiest visualization to grasp, and helped participants see what types of errors the two models had: ``\textit{What's really positive is it's very easy to understand}''  (PID13); ``\textit{It does give a quick reality check in terms of the actuals versus the predicted}'' (PID8).  

Ratings for the bar chart \textbf{V4} ranged from ``neutral'' to ``strongly unhelpful,'' with all participants remarking that the chart did not seem visually distinguishable between the two models, so it was not clear what the chart was trying to communicate: ``\textit{The shape of the curves are similar, it's just not jumping out at me what conclusion I'm supposed to take from it}'' (PID11).  However, it was noted by PID9 that, for a different scenario, the chart could be helpful if the distribution of errors was very different: ``\textit{Overall shape was the same and the absolute errors were close.  Probably wasn't as useful in the exercise. But there are probably circumstances where it might be helpful as a more differentiating figure.}'' 

We believe that this illuminates a practical consideration for implementing the guidelines: depending on the dataset and the model predictions, visualizations produced by data scientists may not tell an obvious story.  These ambiguous visualizations should not be included in a presentation, as they might be taxing to interpret and could even mislead the audience towards reaching for a false conclusion.

\subsection{Learning Outcomes from Using the Guidelines}

In every circumstance, SMEs found our additional information about the models and explanations for why a model performed poorly to be helpful when choosing between the two models (KNN and XGB). PID8 found the additional information helped him assess potential risks: ``[In the second presentation] \textit{I can weigh each of these different parameters against everything and try to take more calculated risk. But with the first presentation, I'm really not sure because I'm just dependent on what limited information is provided and if it clicks, it clicks. If it doesn't, then I'm going to lose everything.}"

As PID8 explained to us -- when there is less clarity, it is better to be able to understand where the models perform well and perform poorly, so that risk can be calculated, rather than taking a leap of faith. PID9 similarly told us that ``\textit{a more nuanced presentation can give confidence.}'' These findings demonstrate the value of guidelines \textbf{\hlgfour{G4}, \hlgfour{G5}, \hlgfour{G6}}: weaving outliers, strengths, and weaknesses into the model presentation improves the SME's ability to use their own expertise in their assessment of a model's performance.

All SMEs also provided positive feedback on the use of visualization to communicate model performance rather than simple text or metrics. Annotated examples in the visualizations (\textbf{\hlgnine{G1}}) were recounted by SMEs as particularly helpful in providing quick insights and conclusions to be drawn.
PID11 and PID13 both noted that they wanted the ability to pore over the visualizations outside of our presentation to fully digest all of the information:
``\textit{I'm often someone who doesn't react to something initially on a screen\ldots And when it's on the screen and I'm listening to someone present I'm back and forth between listening to the person talking and then looking at the slide\ldots I'm not digesting the visuals}" (PID11). SMEs supported the usefulness of providing a pre-read document (\hlgeight{G8}) containing the visualizations shown in our presentation. 
We believe this also highlights the need for more interactive tools -- we discuss further in Section~\ref{sec:modalities}.

Overall, the feedback received from the demonstration of our guidelines during follow-up interviews was positive. We found that  \textbf{D1}-\textbf{D5} were all either helpful or neutral, and provide some evidence that guidelines \hlgfour{G5}, \hlgfour{G6}, \hlgeight{G7}, \hlgeight{G8}, and \hlgeight{G9} can be partially addressed by data scientists with these easy-to-implement additions. 
SMEs generally found the guideline-improved presentation to be helpful enough that the guidelines have been implemented within their organization. We discuss the institutional impact of our guidelines in Section~\ref{sec:novartis}, and limitations of this follow-up study in Section~\ref{sec:limitations}.

\section{Discussion}
\label{sec:discussion}

\subsection{Institutional impact: Guidelines in Practice}
\label{sec:novartis}

The guidelines presented in this work were developed in collaboration with a large pharmaceutical organization.  The conducted interviews have highlighted not only that multi-disciplinary communication remains a substantial barrier to communication, but also that the effectiveness of collaboration is often overestimated.
Additionally, we found that familiarity with advanced visualization techniques varies widely across users and groups, even if they are part of the same organization.
This finding has highlighted the benefits gained by providing additional context to presentations, and slowing the pace of discussions by not just providing time and opportunities for SMEs to ask questions, but by encouraging them to do so.

Our industry partner has already begun to adopt these guidelines in order to address and mitigate the challenges presented in this work.  
These teams are high-performing, multi-disciplinary groups that make regular use of ML models to address domain science questions.
Although all collaborators have responded positively to these new guidelines, it is not possible to implement the entire suite of recommendations simultaneously.
After in-depth discussions with SMEs, new visualizations will be introduced slowly to allow fluency in one visualization type at a time.

It should be noted that introducing interactivity must be done with care.  An initial attempt at providing interactive visualization techniques to the SMEs required multiple steps: the SMEs needed to first adapt to new visualization methods, new interaction techniques, and new data.  
Only after providing an interactive system that better matched their fluency with visualization techniques was the system adopted and used by the SMEs.
Overall, implementation of these guidelines has resulted in a noticeable change in group dynamics: SMEs are more engaged, asking more questions, and are more positive about the collaboration.

\subsection{Presentation Modalities}
\label{sec:modalities}
There are many modalities for presenting and communicating model performance. In this paper, we evaluated the use of our visualization guidelines in a PowerPoint presentation, however, SMEs told us in our interview study (Section~\ref{sec:interviews}) they would also find value in interactive presentations. Interestingly, when we first brought up an ``interactive presentation,'' none of the SMEs we interviewed knew precisely what we meant. Once the interviewer explained the capabilities for an interactive system to communicate model performance, SMEs were immediately drawn to the idea that they could test various inputs to outputs themselves. One of the SMEs we interviewed mentioned that during previous model performance presentations, she had found herself wondering ``\textit{what if we tested another input instead?}'' However, she felt this line of communication could be disruptive, and did not think it would be possible to see this type of interaction on-demand.

Of course, there is an indisputable trade-off cost to building interactive tools and interfaces for model performance. There is continued research being done to support effective and accessible interaction for web-based tools\cite{sievert2020interactive}, computational notebooks\cite{wu:2020:b2}, reusable ML interfaces\cite{bauerle2022symphony}, and responsive visualization tools\cite{suresh:2022:intuitively} that are practical for data scientists to use. Adding interaction also introduces costs for the audience (\emph{e.g.}, false discoveries) during a presentation\cite{lam2008framework}. Further, adding interaction could be overwhelming or ineffective for an audience if they are unfamiliar with common visualization techniques (see Section~\ref{sec:novartis}). To this end, we suggest introducing interaction only after SMEs show comfort and understanding of static visualizations, \emph{e.g.}, those we demonstrated in Section~\ref{sec:demonstration}.  

Future work should consider the integration (and effects) of interaction into mainstream data science tools, such as pandas and sklearn. Computational notebooks seem to balance interactivity with the costs of building interactive visualizations, and are historically popular with data scientists\cite{alspaugh2018futzing}. A handful of data scientists we interviewed mentioned they design computational notebooks with SMEs in mind: inputs to the model can be modified on-demand, certain cells can be collapsed or left open depending on who is being presented to, and so on. As part of our contributions in this work, we designed and deployed an interactive Observable notebook that has a variety of static and interactive visualizations inspired by our interview findings. 

\begin{figure} 
    \centering
    \includegraphics[width=0.99\linewidth]{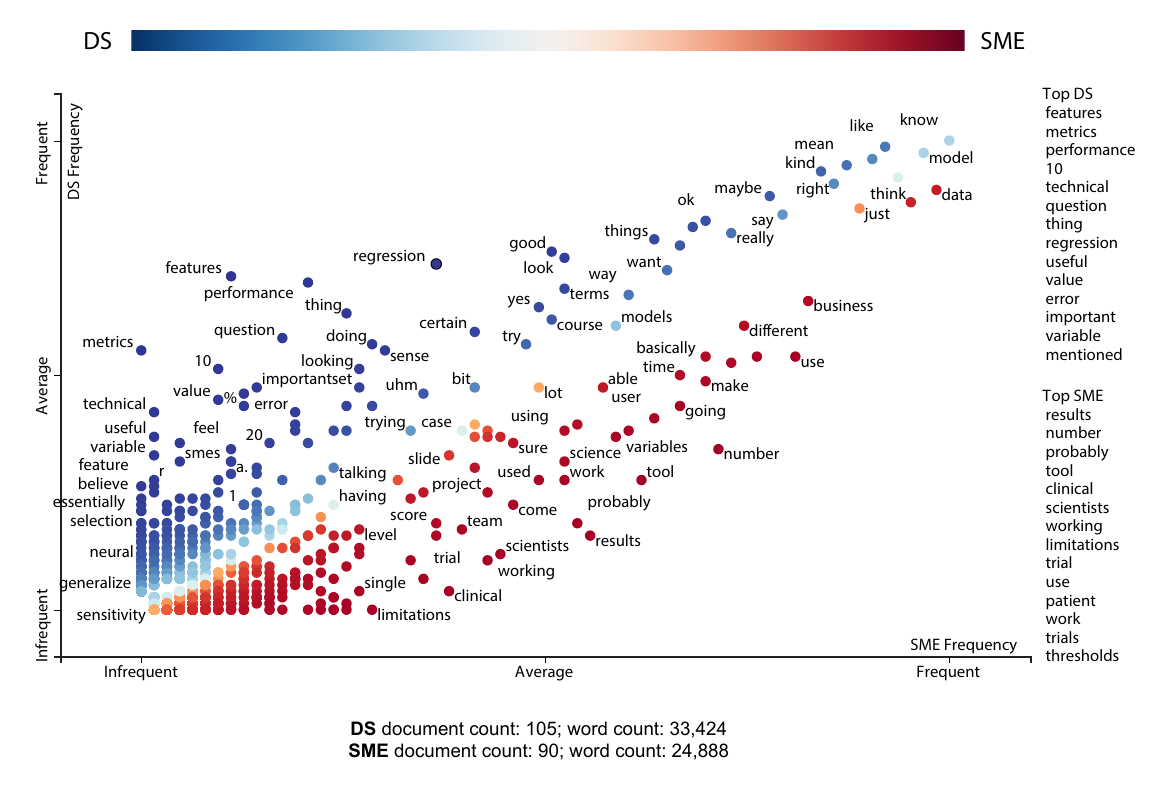} 
    \caption{
    Results from using Scattertext\cite{kessler2017scattertext} on our interview documents. The x-axis represents the ranked word frequency spoken by SMEs (points shown in red), the y-axis represents ranked word frequency spoken by data scientists (points shown in blue), and words shown on the diagonal (colored closer to white) are related to both DS and SMEs. The ``Top DS'' and ``Top SME'' columns show the terms most related to each respective participant group.
    See Section~\ref{sec:interviews:methodology:computation} for more details.
    }
    \label{fig:quantitative_results}
\end{figure}

\subsection{Discriminative Language Analysis}
\label{sec:interviews:methodology:computation}
In addition to our thematic analysis in Sections~\ref{sec:interviews} and~\ref{sec:findings}, we also performed a discriminative analysis between classes using Scattertext\cite{kessler2017scattertext} on our interviews.
Scattertext is a tool that specializes in visualizing linguistic variation between document categories to investigate the differences between the language of the two groups.
The Scattertext plot of the processed interview data can be found in Figure~\ref{fig:quantitative_results}.

The goal of this disciminative analysis was to determine if the two groups were emphasizing different aspects of model communication, or using different terminology to describe the same concepts. 
Ultimately, we can begin to understand the vocabulary data scientists should use for broad audiences in presentations, as well as what might be relevant to add to a pre-read document suggested by \hlgeight{G8}.
The Scattertext results in Figure~\ref{fig:quantitative_results} contains evidence that the two groups did emphasize different aspects of model communication.
Words spoken by data scientists and SMEs with similar frequency tended to be inherent to the collaboration, such as ``models,'' ``data,'' and ``explain.''
Words spoken more frequently by SMEs included terms that related to their own domain: ``clinical,'' ``trial,'' ``patient,'' ``user,'' and ``business,''
illustrating that SMEs tend to ground the discussion of a model within their expertise.
On the other hand, data scientists were more likely to use words like ``features,'' ``regression,'' ``metrics,'' and ``technical,''
demonstrating their propensity to discuss models in a more domain agnostic manner and utilize their own familiar terminology or assessment strategies.
These findings supply further credence that data scientists could work to ground their explanations within the SMEs' domain expertise in order to increase context and comfort (\hlgContext{G9}).
Combining word frequencies with the interview text yields insight into repeatedly spoken words that are used differently by the two groups.
SMEs tended to use ``results'' holistically to describe the performance, limitations, and weaknesses of the model, while data scientists used more specific terminology to describe model performance.
SMEs used ``value'' in the traditional sense of the word (\emph{e.g.}, ``\textit{the model's value in my workplace}''), while data scientists used ``value'' as a prediction or numerical value -- this can lead to confusion across the two groups.
Data scientists often used ``error'' to identify which error SMEs care about, whereas SMEs did not necessarily translate error to the overall goodness of the model.
These slight differences in semantics should be kept in mind by data scientists when delivering presentations to SMEs, and can be included in a pre-read document or appendix slide when appropriate (\hlgContext{G8}).

\subsection{Limitations \& Future Work}
\label{sec:limitations}
The requirements gathered during our interview study, in addition to the guidelines demonstrated and tested in our follow-up study, represent only a subset of problems and solutions from professionals collaborating within the same corporation.  While we recognize that this limits the generalizability of our findings, we believe that it was necessary to restrict the data scientists and SMEs to be in the same organization, in order to get both sides of the same story. The scope of our studies is also narrowed to both the high funding received towards AI/ML at this corporation, and the high familiarity of data science by our interview participants. Moreover, our interviewees are highly educated professionals (4 PhD, 8 MS, 1 BS) who are considered experts in their field.

Our interview study did not attempt to quantify whether the challenges faced by our data scientists and SMEs are common across a variety of domains, skill levels, and corporations. In prior literature, studies have been done to understand AI/ML industry practices across a multitude of domains~\cite{hong2020human}, in addition to sectors outside of `Big Tech'~\cite{hopkins2021machine}. In contrast, our study aims to characterize \textit{how} model performance is assessed and communicated (\emph{e.g.}, presentation styles, language used, metrics, visualizations). Our goal was to identify communication bottlenecks that hinder model usage, and suggest easy-to-implement guidelines that can alleviate these issues through visual mediums. 

Our study is focused on predictive model performance communication, as opposed to general AI/ML interpretability. While it is possible that the methods suggested by our paper could be mapped to communicating any predictive model, we did not ask interview participants any specific questions related to other types of models. Previous work has investigated challenges related to workflows on ``blackbox'' AI models between data scientists and stakeholders (\emph{e.g.},\cite{passi2018trust}), however, we observe a lack of prescriptive guidelines for data scientists to consider when communicating their models. Future work should be done to assess the validity of (or extend) our guidelines to other AI/ML models.

Finally, although a motivation for this study was to help increase the number of models deployed in practice, we could not validate whether our guidelines address this gap. Future work should be done to closely follow a model's development and how it is communicated - aided by our guidelines. Fieldwork by Passi and Jackson\cite{passi2018trust} begins to tackle this critical problem. 
It should similarly be documented and evaluated in research whether our suggested guidelines impact a model's eventual adoption in practice.

\section{Conclusion}
\label{sec:conclusion}
We present guidelines for communicating predictive model performance within a large organization.  Based on interviews with both data scientists and subject matter experts, we identify common gaps in communication and suggest broadly applicable solutions for data scientists to use in presenting their results to SMEs. We present and elicit feedback for a demonstration of our guidelines in a data science presentation given to SMEs. Our results indicate that our guidelines are helpful in producing a better understanding of the nuances of a model's performance -- including its strengths, weaknesses, and trade-offs -- beyond text, numbers, and commonly used model metrics.

\section*{Acknowledgments}
  This work was supported by National Science Foundation grants IIS1452977, OAC-1940175, OAC-1939945, OAC-2118201, NRT-2021874, and DOD grant HQ0860-20-C-7137. We express our sincere gratitude to our collaborators at Novartis for their time and participation in our study. We thank the reviewers for their valuable feedback and suggestions in improving the quality of our manuscript.


\ifCLASSOPTIONcaptionsoff
  \newpage
\fi



%


\bibliographystyle{IEEEtran}
\bibliography{IEEEabrv,main}

%
\begin{IEEEbiography}[{\includegraphics[width=1in,height=1.25in,clip,keepaspectratio]{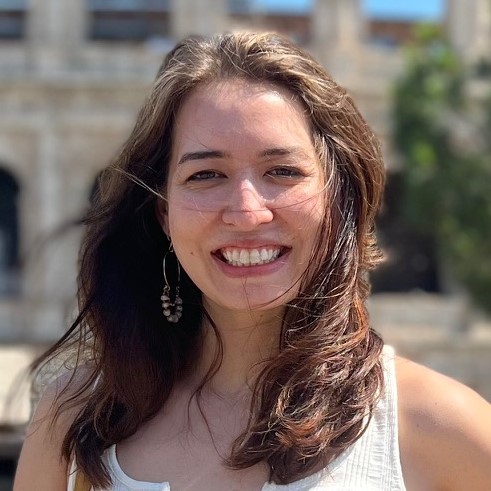}}]{Ashley Suh}
received her MS in Computer Science from Tufts University where she is currently pursuing a PhD. Her research interests include information visualization, visual analytics, and human-centered AI.
\end{IEEEbiography}

\begin{IEEEbiography}[{\includegraphics[width=1in,height=1.25in,clip,keepaspectratio]{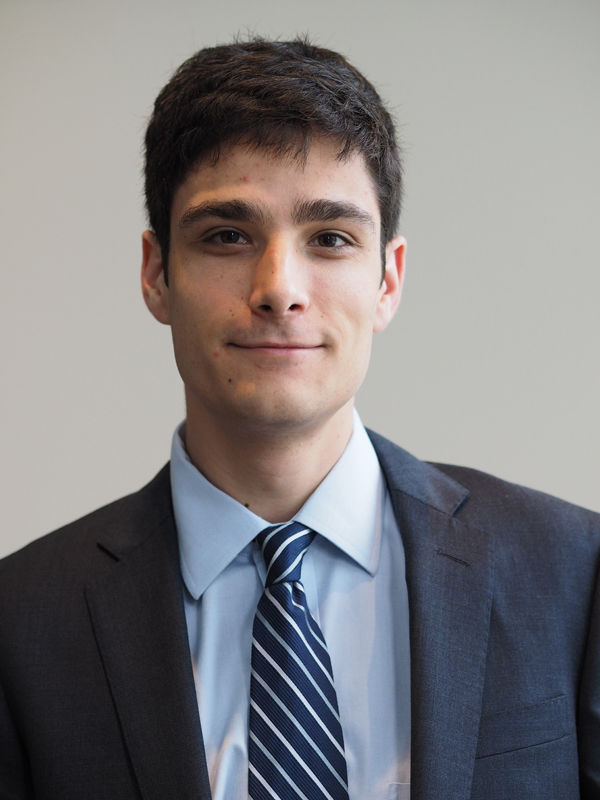}}]{Gabriel Appleby} received his MA degree in Computer Science at Tufts University where he is currently working towards his PhD. His research spans the fields of data visualization, visual analytics, and machine learning.
\end{IEEEbiography}

\begin{IEEEbiography}[{\includegraphics[width=1in,height=1.25in,clip,keepaspectratio]{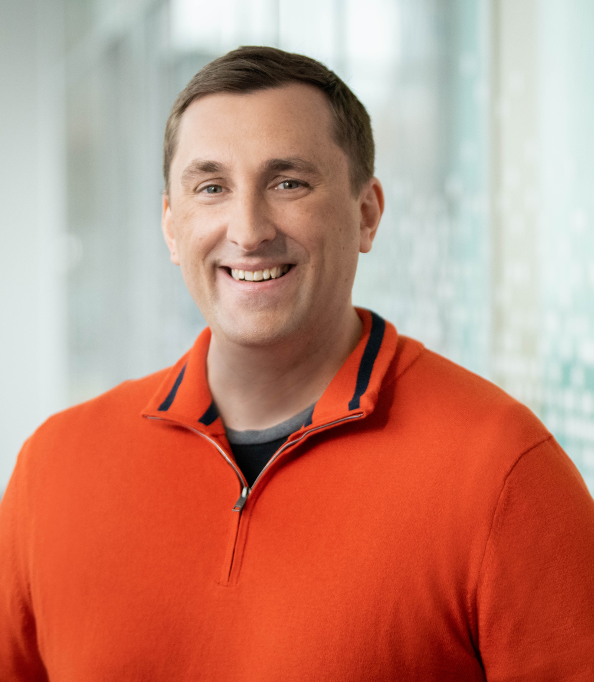}}]{Erik W Anderson}
received his PhD (2011) in Scientific Computing from the University of Utah, USA.  He was a senior scientist for Electrical Geodesics, Inc (EGI) until 2017 and then worked in R\&D at Philips Neuro until 2020. Since 2020 he has been the Head of the Visualization and Visual Analytics group and Novartis Inc’s AI Innovation Center in Cambridge, MA USA.  His interests include high-dimensional visualization, multi-modal modeling and visualization, and biomedical image visualization.
\end{IEEEbiography}

\begin{IEEEbiography}[{\includegraphics[width=1in,height=1.25in,clip,keepaspectratio]{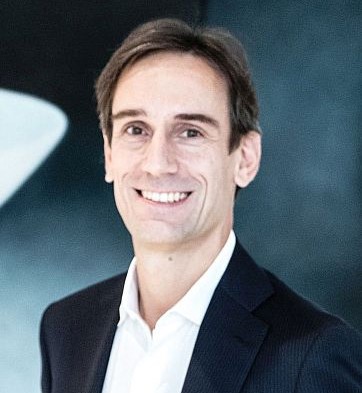}}]{Luca Finelli} received his PhD in Neuroscience from the Swiss Federal Institute of Technology (ETH) Z\"{u}rich. He is the Head of Insights Strategy \& Design for the Data Science and AI unit in Novartis Digital and Program and System Owner of Nerve Live, a data and analytics platform harnessing past and present operational data across Novartis. His research interests span theoretical physics to big data, analytics, and medicine.
\end{IEEEbiography}

\begin{IEEEbiography}[{\includegraphics[width=1in,height=1.25in,clip,keepaspectratio]{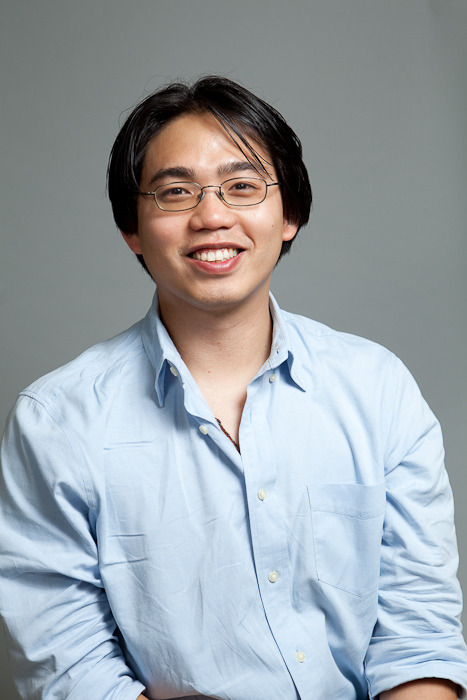}}]{Remco Chang} received his PhD in computer science from the University of North Carolina Charlotte. He is an associate professor in computer science with Tufts University. His research interests include visual analytics, information visualization, human computer interaction, and databases.
\end{IEEEbiography}

\begin{IEEEbiography}[{\includegraphics[width=1in,height=1.25in,clip,keepaspectratio]{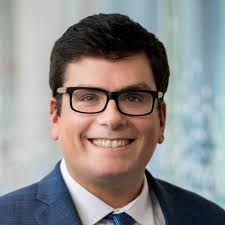}}]{Dylan Cashman} received his PhD in Computer Science from Tufts University. He received a bachelor of science in Mathematics from Brown University. Since 2020 he is a senior expert in data science and advanced visual analytics in the Insights, Strategies, and Design group at Novartis Pharmaceuticals. His research interests include visualization for data science and interactive machine learning.
\end{IEEEbiography}




\end{document}